\documentclass[11pt]{article}

\usepackage{amssymb}
\usepackage{amsmath}
\usepackage{graphicx}
\usepackage{hep}

\newcommand{\hzero}{\ensuremath{\PHiggslightzero}} %light neutral Higgs
\newcommand{\Hzero}{\ensuremath{\PHiggsheavyzero}} %heavy neutral Higgs
\newcommand{\Azero}{\ensuremath{\PHiggspszero}} % CP-odd neutral Higgs
\newcommand{\Hpm}{\ensuremath{\PHiggspm}} %charged Higgs

\newcommand{\CP}{\ensuremath{\mathcal{C}\mathcal{P}}}

\newcommand{\bsg}{\ensuremath{\mathcal{B}(b \to s \gamma)}}

\newcommand{\sbt}{\sin\beta}

\newcommand{\cbt}{\cos\beta}

\newcommand{\mh}{M_{h^0}}
\newcommand{\mH}{M_{H^0}}

\newcommand{\mA}{M_{A^0}}

\newcommand\pubblock{\rightline{\begin{tabular}{l} \pubnumber\\
     \pubdate\\  \end{tabular}}}
\newcommand\pubnumber{UB-ECM-PF-08/21}
\newcommand\pubdate{January 2009 }

\def\beq{\begin{eqnarray}}    %%%  begequation/eqnarray
\def\eeq{\end{eqnarray}}      %%%  endequation/eqnarray
\newcommand{\mysection}[1]{\section{#1}
\renewcommand{\theequation}{\thesection.\arabic{equation}}
\setcounter{equation}{0}}

\newcommand{\mysubsection}[1]{\subsection{#1}
\renewcommand{\theequation}{\thesubsection.\arabic{equation}}
\setcounter{equation}{0}}

\begin{document}
\pubblock

\begin{center}
{\large \textsc{Higgs boson pair production through gauge boson
fusion \\ at linear colliders within the general 2HDM}} \vskip
2mm
 \vskip 8mm

\textbf{Robert N. Hodgkinson}$^{a,b}$, \textbf{David
L\'opez-Val}$^{a,c}$, \textbf{Joan Sol\`{a}}$^{a,c}$

\vskip0.5cm

$^{a}$ High Energy Physics Group, Dept. ECM, Univ. de Barcelona\\
Av. Diagonal 647, E-08028 Barcelona, Catalonia, Spain

$^{b}$ School of Physics and Astronomy, University of Manchester,
Manchester, England.

$^{c}$ Institut de Ci\`{e}ncies del Cosmos, UB, Barcelona \\

E-mails: robert.hodgkinson@postgrad.manchester.ac.uk,
dlopez@ecm.ub.es, sola@ecm.ub.es. \vskip2mm

\end{center}
\vskip 15mm
%%%%%%%%%%%%%%%%%%%%%%%%%%%%%%%%%%%%%%%%%%%%%%%%%%%%%%%%%%%%%
\begin{quotation}
\noindent {\large\it \underline{Abstract}}.\ \ Inclusive Higgs boson
pair production through the mechanism of gauge boson fusion
$\APelectron\Pelectron \to V^*V^*\to h\,h + X$ ($V=W^{\pm},Z$) in
the general Two-Higgs-Doublet Model (2HDM), with $h =
\PHiggslightzero,\PHiggsheavyzero,\PHiggspszero, \PHiggs^\pm$, is
analyzed at $\mathcal{O}(\alpha^4_{ew})$ in the linear colliders ILC
and CLIC. This kind of processes is highly sensitive to the
trilinear (3H) Higgs boson self-interactions and hence can be a true
keystone in the reconstruction of the Higgs potential. For example,
in the ILC at $1\,$ TeV, the most favorable scenarios yield
cross-sections up to roughly $1\, \picobarn$, thus entailing
$10^{5}$ events per $100\,\invfb$ of integrated luminosity, whilst
remaining fully consistent with the perturbativity and unitarity
bounds on the 3H couplings, the electroweak precision data and the
constraints from $\bsg$. Comparing with other competing mechanisms,
we conclude that the Higgs boson-pair events could be the dominant
signature for Higgs-boson production in the $\TeV$-class linear
colliders for a wide region of the 2HDM parameter space, with no
counterpart in the Minimal Supersymmetric Standard Model (MSSM).
Owing to the extremely clean environment of these colliders,
inclusive 2H events should allow a comfortable tagging and might
therefore open privileged new vistas into the structure of the Higgs
potential.
\end{quotation}
\vskip 8mm

%PACS numbers:\ {95.36.+x, 04.62.+v, 11.10.Hi}

\vskip 6mm

\newpage

 \noindent \mysection{Introduction}
 \label{Introduction}

\hspace{0.6cm}For more than 40 years, the Standard Model (SM) of
Strong and Electroweak interactions has furnished a successful arena
in which to describe the physics of Elementary Particles. In spite
of its formidable achievements, a number of longstanding challenges
are still to be resolved. Perhaps the most paradigmatic one concerns
the ultimate nature of Electroweak Symmetry Breaking (EWSB). Even
though the Higgs mechanism provides an elegant description of EWSB
within a perturbative quantum field theory framework, one is forced
to postulate the existence of (at least) one scalar ($J^{P}=0^{+}$)
fundamental building-block of Nature, whose experimental
confirmation is conspicuously missing for the time being.
Nevertheless, with the recent start-up of the LHC operations at
CERN, the prospects for the discovery of the Higgs boson are
extremely encouraging and the curtains may soon be drawn back to
reveal this final player in the story of the SM. In fact, the LHC
will be able to amply sweep the natural SM Higgs mass range
(spanning from a few hundred GeV up to $\sim 1\,$TeV) and, in the
most favorable scenarios, it could produce a copious number of Higgs
boson events\,\cite{HWG08}.%\jump

Despite these optimistic prospects for the discovery of the SM Higgs
boson, the upcoming LHC data might well reveal that the ultimate
origin of EWSB is  grounded somewhere beyond the minimal assumptions
of the SM. For instance, a number of well-motivated (perturbative)
extensions of the SM entail a two-Higgs-doublet structure, such as
the Minimal Supersymmetric Standard Model (MSSM) \cite{susy} or the
general (unconstrained) Two-Higgs-Doublet Model (2HDM)
\cite{hunter}. From the beginning, however, the quest for
experimental signatures of Higgs boson physics beyond the SM
concentrated its efforts mainly on the search for supersymmetric
(SUSY) Higgs bosons \,\cite{susy,hunter}-- see \cite{Frank06} for
more recent developments, and \cite{Djouadi08} for fresh reviews of
this subject. At the same time, very detailed investigations were
undertaken in the past on the SUSY quantum effects on the gauge
boson masses\,\cite{Masses} (cf. \cite{art} for the current state of
the art) and on the $Z$-boson and top quark widths\,\cite{Widths},
in which virtual Higgs bosons may also play a significant role.
Later on, outstanding new sources of quantum corrections were
identified in processes involving Higgs bosons, quarks and/or
squarks as external particles. The possible new effects stemmed from
the enhancement capabilities associated to the supersymmetric Yukawa
couplings between Higgs bosons and quarks or between quarks, squarks
and chargino-neutralinos. In a variety of quite different processes,
the potential effects were shown to be truly spectacular, see e.g.
\,\cite{Coarasa:1996qa,GHS0298,CarenaGNW,BGGS02}.  More recently, it
has been shown that such an enhancement also applies to the
loop-generated Yukawa couplings of additional gauge-singlet Higgs
bosons in minimal extensions of the MSSM \cite{HP1}.

So far so good, but what about the \textit{non}-SUSY Higgs boson
extensions of the SM? Remarkably, the crucial novelty here is that
the most relevant effects could have an origin fundamentally
different from the Yukawa couplings linked to the traditional SUSY
sources of enhancement. This possibility has recently been
illustrated for the general 2HDM in\,\cite{previous} (see also
\cite{Bejar06} in a different domain). In this letter, we dwell on
another compelling (and complementary) example of this fact, which
we encounter once more in the physics of the Higgs boson
self-couplings at linear colliders. Assuming that the LHC uncovers a
neutral Higgs boson, a critical issue will be to discern whether
this particle is compatible with the SM and/or any of its extensions
and, in the latter case, to which of these extensions it belongs.
The next generation of $\TeV$-class linear colliders (based on both
$\APelectron \Pelectron$ and $\Pphoton\Pphoton$ collisions), such as
the ILC and CLIC projects\,\cite{ILC-CLIC}, will be invaluable in
answering this fundamental question. Owing to its extremely clean
environment, a linear collider should allow for precise measurements
of: i) the Higgs boson mass (or masses, if more than one); ii) the
couplings of the Higgs bosons to quarks, leptons and gauge bosons;
and iii) the Higgs boson self-couplings mentioned above, i.e. the
trilinear (3H) and quartic (4H) Higgs boson self-interactions. In a
nutshell, it should allow us a keen insight into the
physics lying beneath the EWSB mechanism and, ultimately, to
reconstruct the Higgs potential itself.

In this work, we provide some complementary clues to this
reconstruction process: specifically, we focus on the effects of the
trilinear Higgs boson couplings on the inclusive production of Higgs
boson pairs induced by weak gauge boson fusion, i.e.
$\APelectron\Pelectron \to V^*V^*\to h\,h + X$, where $V=W^{\pm},Z$
and $h = \PHiggslightzero,\PHiggsheavyzero,\PHiggspszero,
\PHiggs^\pm$. We show that this mechanism could be the leading Higgs
boson production channel at TeV energies and, if so, it should
furnish an intrinsic and totally unambiguous signal of
non-supersymmetric new physics. Therefore, while the physics of the
top (and bottom) quark\,\cite{Bernreuther08} is the natural
playground for the study of the SM and MSSM Higgs boson
interactions, we find that in the case of \textit{non}-SUSY theories
there are comparable (if not better) opportunities in sectors of the
theory not necessarily related to heavy quark flavors but to the
very structure of the Higgs potential.

%%%%%%%%%%%%%%%%%%%%%%%%%%%%%%%%%%%%%%%%%%%%%%%%%%%%%%%%%%%%%%%
\noindent \mysection{Higgs boson production induced by trilinear
Higgs interactions} \label{sect:3H}
%%%%%%%%%%%%%%%%%%%%%%%%%%%%%%%%%%%%%%%%%%%%%%%%%%%%%%%%%%%%%%%

%\jump
Of cardinal importance is to understand in detail the phenomenology
of the Higgs sector in the context of linear colliders (both within
the SM and its renormalizable generalizations). Quite an effort has
already been devoted to this goal in the literature. Exclusive
double (2H) and multiple Higgs boson production, for instance, has
been comprehensively investigated, although mainly within the MSSM\,
\cite{Djouadi:1992pu,pairmssm,Feng:1996xv}. The exclusive 2H case
consists of the simplest Higgs boson production processes:
\begin{eqnarray}
  \APelectron\Pelectron \to 2\PHiggs\,\ \ \ \ \ (2\PHiggs \equiv \hzero\,\Azero; \Hzero\,\Azero;
\PHiggsplus\PHiggsminus)\,. \label{2H}
\end{eqnarray}
Similarly, the two-body final states $\APelectron\Pelectron \to \PZ\PHiggslight$ and
 $\APelectron\Pelectron \to \PHiggsps\PHiggslight$ {(with $\PHiggslight =
\hzero\, ,\Hzero$)} have been long known to be complementary to each
other in the MSSM\,\cite{Djouadi:1992pu}. Notice that processes with
exclusively two identical neutral Higgs bosons in the final state
cannot proceed at the tree-level (neither in the SM nor in the
MSSM), and at one-loop they have very tiny cross-sections of order
$10^{-5}$ pb \,\cite{previous}. However, sizeable rates of two Higgs
bosons in the final state in an $\APelectron\Pelectron$ collider may
arise from the processes {like (\ref{2H}), whose detection would
signify} an unmistakable observation of physics beyond the SM. {
Nevertheless, let us highlight that none of the exclusive 2H
channels (\ref{2H}) is sensible to the trilinear Higgs boson
couplings at the leading order. Within such pairwise Higgs events,
therefore, probing the structure of the 3H self-interactions would
only be possible through the analysis of the radiative corrections.
Indeed, one needs to include such quantum effects in the exclusive
2H boson processes (\ref{2H}) so as to disentangle the genuine
imprints of a SUSY Higgs sector from a non-SUSY one.} In this
context, a rich literature exists on the one-loop calculation of the
cross-sections for the two-particle final states within the
MSSM\,\footnote{See \cite{mssmloop,Djouadi:1999gv,Fawzy02}, and also
the extensive report \cite{Weiglein:2004hn} and references
therein.}. There are also studies considering radiative corrections
to charged Higgs production in $\APelectron\Pelectron$ collisions
within the 2HDM\,\cite{ghk}, and also on single, double and multiple
Higgs production at the LHC but mostly for the MSSM\,\cite{inlhc}. A
complete $1$-loop analysis of the exclusive 2H channels (\ref{2H})
in the general 2HDM is missing, however, and will be presented
elsewhere\,\cite{paper2}.

{Our main endeavor in this letter is to further investigate the
foreseeable impact of these 3H self-couplings in a class of
processes which critically depend on them already at the tree-level.
This was for instance the case of Ref.\,\cite{previous}, in which}
the triple Higgs boson couplings were probed by performing a
systematic analysis of the exclusive production processes with three
Higgs bosons in the final state. There are three classes of
processes of this kind compatible with $\CP$-conservation, to wit:
\begin{eqnarray}
1)\ \APelectron\Pelectron \to \PHiggsplus\PHiggsminus \PHiggslight
\, ,\ \ \ 2)\ \APelectron\Pelectron \to \PHiggslight \PHiggslight
\Azero\, , \ \ \ 3)\ \APelectron\Pelectron \to \hzero \Hzero
\Azero\,, \ \ \ (\PHiggslight=\hzero,\Hzero,\Azero)\,, \label{3H}
\end{eqnarray}
where, in class 2), we assume that the two neutral Higgs bosons
$\PHiggslight$ must be the same, i.e. the allowed final states are
$(\PHiggslight\, \PHiggslight \Azero)=(\hzero \hzero \Azero)$,
$(\Hzero \Hzero \Azero)$ and $(\Azero \Azero \Azero)$.  The
cross-sections in the 2HDM were shown in \cite{previous} to reach up
to ${\cal O}(0.1)\,\picobarn$, i.e. several orders of magnitude over
the corresponding MSSM predictions\,\cite{Djouadi:1992pu}. Similar
effects have also been recently reported in 2H strahlung processes
of the guise $\APelectron\Pelectron \HepTo \PZ^0\, h\,h$
\cite{arhrib08}, and also in the loop-induced 2H production through
$\Pphoton\Pphoton$ interactions \cite{hollik08}.

It is important to note that, in a $\CP$-conserving theory, all the
3H final states in (\ref{3H}) contain at least one charged or
pseudoscalar Higgs boson, and this has practical implications. In
fact, let us recall that there is an important distinction between
the two basic types of generic 2HDM models\,\cite{hunter}; namely,
whilst light charged Higgs bosons are possible within a type-I 2HDM
-- in which only the $\Phi_2$ field couples to
fermions\,\footnote{Throughout the paper, we use the notation and
conventions of Ref.\,\cite{previous}.} --, in type-II models there
are important constraints on the charged Higgs mass due to
contributions to the flavor-changing neutral-current (FCNC) process
$b\to s\gamma$\,\cite{gamba,pdg08}. The Higgs pseudoscalar $\Azero$
is then also constrained to be relatively heavy $M_A\sim M_{H^\pm}$,
due to the limits on $\delta\rho$ -- see e.g. Eq.\,(2.6) of
Ref.\cite{previous}.

%\jump
Triple Higgs boson couplings may drive different kinds of processes.
The phenomenological impact, however, may seriously depend on
whether the underlying Higgs sector belongs to the MSSM or to the
general 2HDM. In practice, this means that we should be ready to
identify highly distinctive signatures of the underlying model. We
have mentioned above that the trilinear Higgs boson couplings are
involved at the tree-level in processes with three Higgs bosons (3H)
in the final state, see (\ref{3H}). But, in fact, they are also
involved in inclusive processes with two Higgs bosons (indicated as
2HX) in the final state. For instance, the 3H-coupling has been
investigated phenomenologically in TeV-class linear colliders in
\cite{pairmssm,Djouadi:1999gv,Fawzy02} through the double-Higgs
strahlung process $\APelectron\Pelectron \to \PHiggsheavy
\PHiggsheavy \PZ$ or the $\PW\PW$ double-Higgs fusion mechanism
$\APelectron \Pelectron \to \PHiggsplus\PHiggsminus \Pnue\APnue$.
These processes, which include vertices like $\PZ\PZ\PHiggsheavy$,
$\PW\PW\PHiggsheavy$, $\PZ\PZ \PHiggsheavy \PHiggsheavy$,
$\PW\PW\PHiggsheavy\PHiggsheavy$ and $\PHiggsheavy \PHiggsheavy
\PHiggsheavy$, are possible both in the SM and its extensions, such
as the MSSM and the general 2HDM.  Unfortunately, the cross-section
turns out to be rather small  both in the SM and in the MSSM, being
of order $10^{-3}$ pb at most, i.e. equal or less than $1$ fb\,
\cite{Djouadi:1999gv}. Even worse is the situation regarding the 3H
boson production in the MSSM, in which -- except in the case of some
particular resonant configuration -- the typical cross-sections just
border the value $\sim 0.01\ $fb or less \,\cite{Djouadi:1999gv}. In
the latter reference it has been shown that, if the double and
triple Higgs production cross-sections would yield sufficiently high
signal rates, the system of couplings and corresponding
double/triple Higgs production cross-sections could in principle be
solved for all trilinear Higgs self-couplings up to discrete
ambiguities using only these processes. But in practice these
cross-sections are manifestly too small to be all measurable.

In stark contrast with the pessimistic situation in the MSSM, the
unconstrained 2HDM may exhibit quite promising signatures.  A key
point here is the potentially large enhancements that the 3H
couplings may accommodate -- unlike the MSSM case, where such
self-couplings are constrained by SUSY invariance and are all
predicted to be purely gauge \,\cite{hunter,Djouadi08}. {To make it
transparent with a single explicit example, the analytical
expression for the particular
$\PHiggslightzero\PHiggslightzero\PHiggsheavyzero$ self-coupling in
both models reads}\,\footnote{The complete list of trilinear Higgs
boson couplings in the general 2HDM is presented in Table 1 of
Ref.\,\cite{previous}.}

\begin{eqnarray}
C_{\mbox{2HDM}}[\PHiggslightzero\PHiggslightzero\PHiggsheavyzero] &=& -\frac{i\,e\,\cos(\beta-\alpha)}{2\,M_W\sin\theta_W\,
             \sin{2\beta}
             }\,
            \left[(2\,\mh^2+\mH^2)\,\sin{2\alpha}
            -\mA^2\,(3\sin{2\alpha}-\sin{2\beta})\right] \nonumber \\
C_{\mbox{MSSM}}[\PHiggslightzero\PHiggslightzero\PHiggsheavyzero] &=&
\frac{i\,e\,M_Z}{2\,\cos\theta_W\,\sin\theta_W}\,\left[\cos\,2\alpha\,\cos(\alpha+\beta)
- 2\,\sin 2\alpha\, \sin(\alpha+\beta)\right]
 \label{eq:3Hcoup}\,.
\end{eqnarray}
{It is patent from these expressions that the 2HDM coupling can be
enhanced both at low and high $\tan\beta$, and also through the
Higgs boson mass splittings, whereas the MSSM coupling cannot be
enhanced in any way. In practice, the enhancement possibilities of
the 2HDM couplings will be partially tamed by the aforementioned set
of phenomenological and theoretical restrictions.}

\begin{figure}[t]
\begin{center}
\begin{tabular}{c}
 \includegraphics[scale=0.9]{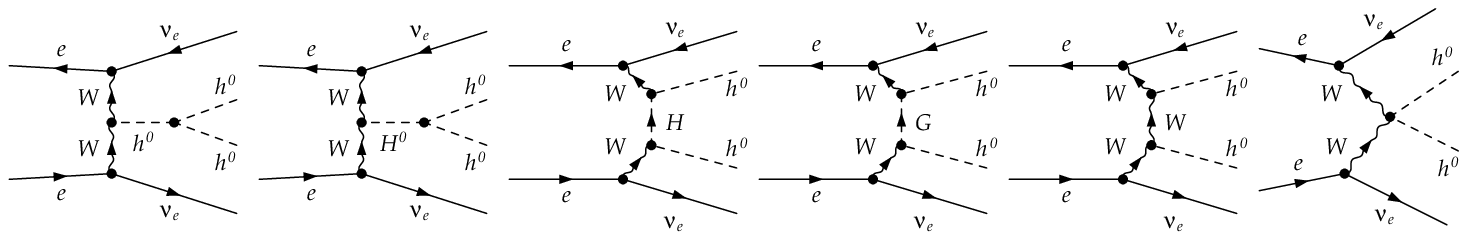} \\ \quad \\
  \includegraphics[scale=0.9]{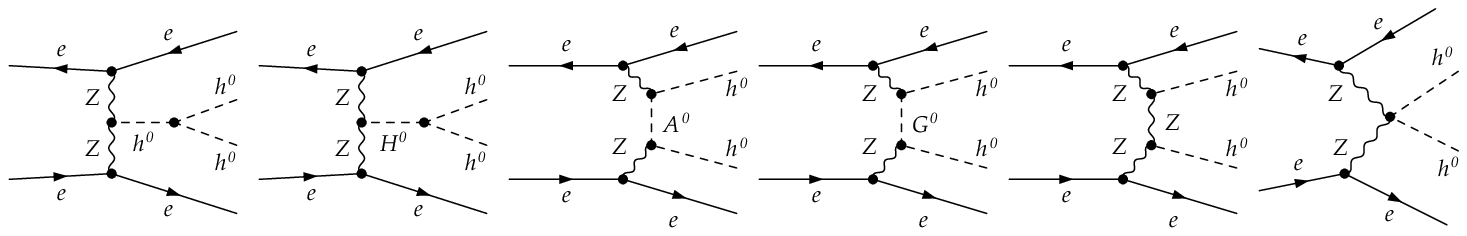}
\end{tabular}
\end{center}
\caption{\footnotesize{Dominant Feynman diagrams contributing to
$\APelectron\Pelectron \to V^*\,V^*\,\to\,
\PHiggslightzero\PHiggslightzero \,+ X$ at high energy -- with
$X=(\APelectron,\ \Pelectron\,;\, \bar{\nu_e},\ \nu_e)$. Notice the
presence of trilinear couplings of the sort
$\PHiggslightzero\PHiggslightzero\PHiggslightzero$ and
$\PHiggsheavyzero\PHiggslightzero\PHiggslightzero$. These Feynman
diagrams have been obtained with the help of the computational
package \emph{FeynArts} \cite{feynarts}. There are many other
diagrams contributing to the same final state, which are not of
gauge fusion type and are not shown in this figure. Our
cross-section calculation, however, includes the complete set.}}
\label{Feynman_diagram}
\end{figure}

In the next section, we analyze the \emph{inclusive} Higgs
boson-pair production at linear colliders within the general 2HDM,
\begin{eqnarray}
\APelectron\Pelectron \to h\, h \,+ X\ \ \ \ \ \ (h =
\PHiggslightzero,\PHiggsheavyzero,\PHiggspszero,\PHiggs^{\pm})\,,\label{2HX}
\end{eqnarray}
\noindent where $X=(\APelectron,\ \Pelectron\,;\, \bar{\nu_e},\
\nu_e)$. At high energies ($\sim$TeV) the vector boson fusion
diagrams of the kind displayed in Fig.~\ref{Feynman_diagram}
constitute the dominant mechanism. Therefore, in practice the bulk
mechanism of (\ref{2HX}) is $\ \APelectron\Pelectron \to V^*V^*\to
h\, h \,+ X$. Notice that the fusion mechanism may trigger either
one or no Higgs boson as a virtual intermediate state. There is also
the possibility that a single real Higgs boson is produced on-shell
and subsequently decays into two Higgs bosons of smaller mass. In
this respect, let us recall that, in the SM, the mechanism of single
Higgs boson production via gauge boson fusion in
$\APelectron\Pelectron$ collisions was already considered long
ago\,\cite{JonesPetcov79} and was shown to be dominant with respect
to the annihilation channels at high energy. Still, the
cross-sections are very small in the SM (of order $1-10$ fb) for
masses $M_H\sim 100$ GeV and they were computed at that time for the
``future'' LEP energies. Some enhancement can be achieved for
charged Higgs production in extended Higgs models\,\cite{GS81}.
However, none of the last two sorts of processes involve the Higgs
boson self-couplings.

In the following, we concentrate on the computation of the
cross-section for the processes (\ref{2HX}), which do involve the
3H-couplings in some of their Feynman diagrams, see
Fig.~\ref{Feynman_diagram}. It turns out that these specific
diagrams are responsible for the bulk of the cross-sections under
the most favorable conditions for these processes. We will perform
the calculation in the general 2HDM model and shall take the same
set of phenomenological restrictions used in Ref.\,\cite{previous},
to which we also refer the reader for more details on the relevant
pieces of the interaction Lagrangian.

%
%%%%%%%%%%%%%%%%%%%%%%%%%%%%%%%%%%%%%%%%%%%%%%%%%%%%%%%%%%%%%%%
\noindent \mysection{Double Higgs boson production from weak gauge
boson fusion} \label{sect:2HXintro}
%%%%%%%%%%%%%%%%%%%%%%%%%%%%%%%%%%%%%%%%%%%%%%%%%%%%%%%%%%%%%%%

In contrast to the simple 2H channels (\ref{2H}), the class of
triple Higgs boson processes (\ref{3H}) and the inclusive 2H
processes (\ref{2HX}) are both directly sensitive to the trilinear
self-interactions, which implies both a source of potential
enhancement and a possible strategy to measure such couplings. In
Ref.~\cite{previous} it was shown that, for Higgs boson masses
$\gtrsim 100\,GeV$, the production cross-sections corresponding to
the 3H processes (\ref{3H}) could be remarkably high in the general
2HDM, lying typically above the inclusive 2H cross-sections
(\ref{2H}) at center-of-mass (CM) energies $\gtrsim1$~TeV. This
feature, which is impossible in the MSSM, is brought about by the
enhanced Higgs boson self-couplings involved in the triple Higgs
production mechanism (\ref{3H}).

In a similar way, when we move to the 2HX processes (\ref{2HX}) and
consider higher and higher energy (typically at the $\sim$ TeV range
of linear colliders), several opportunities for significant
enhancement may appear. The leading mechanism behind these processes
is the weak gauge-boson fusion mechanism
\begin{eqnarray}
\APelectron\Pelectron \to V^*V^*\to h\, h \,+ X\ \ \ \ \ \
(V=W^{\pm}\, Z;\ \ h =
\PHiggslightzero,\PHiggsheavyzero,\PHiggspszero,\PHiggs^{\pm})\label{2HXF}
\end{eqnarray}
(cf. Fig.~\ref{Feynman_diagram} for the case $h=\PHiggslightzero$).
As a result, the three main sources of enhancement here are the
following: i) first, the $t$-channel gauge boson fusion is not so
severely damped at high energies as compared to s-channel
annihilation; ii) second, the triple Higgs vertex is involved in
some of the gauge boson fusion channels (see
Fig.\,\ref{Feynman_diagram}, first and second diagrams from the left
in either row); and, finally, iii) in some cases, the virtual Higgs
boson produced in the last sort of diagrams may not be too far away
from the resonance and, in such circumstance, the 2HX final state
can be more copiously produced. In particular, if the Higgs boson
$H^0$ is heavy enough, the resonant decay $H^0\to h^0 h^0$ will be
kinematically allowed. Although this decay mode is also possible
within the MSSM, the production of the initial $H^0$ is mostly
suppressed in this model, since its gauge couplings are known to be
complementary to those of the SM-like $h^0$.  The rate for this
process in the MSSM is therefore not competitive with the 2HDM one.
The upshot is that some of the gauge boson fusion processes
(\ref{2HXF}) can become fully competitive, if not the leading
mechanism of Higgs boson production, at high energy in the general
2HDM.

Let us emphasize that the cross-sections for the processes
(\ref{2HXF}) grow up to very high values of the CM energy
$\sqrt{S}$. This is possible because, for the fusion diagrams, the
weak gauge bosons $V$ can be quasi-real and hence have virtual
momenta well-below the CM energy of the process, which may satisfy
${S}\gg M^2_V$ -- the rest of the energy being carried away by the
concomitant lepton final states $X=(\APelectron,\ \Pelectron\,;\,
\bar{\nu_e},\ \nu_e)$. Therefore, while the exclusive 2H and 3H
production cross-sections are expected to decay irremissibly as
$\sim 1/S$ owing to the $Z$-boson propagator that mediates the
s-channel diagrams, the energy behavior of the gauge fusion
mechanism for 2HX production is quite different. It is actually
reminiscent of the cross-section for two-photon processes
$\APelectron\Pelectron \to \gamma^*\,\gamma^*\to Y+
\APelectron\Pelectron$, which in the asymptotic limit goes roughly
as $\sim(\alpha^4/M^2)\,\ln^2(S/m_e^2)\,\ln^n(S/M^2)$, where $M$ is
the threshold mass of the produced final state $Y$ and the number
$n\geq 1$ depends on the high energy behavior of
$\sigma(\gamma\gamma\to Y)$ \footnote{The corresponding analysis of
the single Higgs production through photon-photon fusion,
$\APelectron\Pelectron \to \Pphoton^*\Pphoton^* \to h + X$ within
the general 2HDM will be discussed in a forthcoming publication
\cite{future}.}. In our case, the situation is a bit more
complicated because we have massive gauge bosons $V$ (rather than
photons), and moreover the diagrams with triple Higgs vertex contain
an intermediate virtual Higgs state emerging from the $VV$-fusion.
Thus, the dependence of the cross-sections (\ref{2HX}) with the
threshold mass of the Higgs boson pair is not so simple, but as in
the two-photon case the cross-sections are expected to raise
logarithmically with the energy, rather than decaying quadratically
with it. We shall confirm these expectations with the numerical
analysis in the next section.

The following comment is also in order. Even though the results on
the inclusive Higgs boson-pair production (\ref{2HX}) are
overwhelmingly dominated by the gauge boson fusion mechanism
(\ref{2HXF}), we point out that we have computed the cross-section
of the processes (\ref{2HX}) with full generality, i.e. by including
all the diagrams at order $\mathcal{O}(\alpha^4_{ew})$. This is
actually necessary in order to insure the gauge invariance of the
overall result. Some subsets of diagrams are nonetheless completely
irrelevant, such as e.g. those $Z$-mediated amplitudes where the
Higgs bosons are radiated off the lepton legs. However, there are
other more subtle subsets. In particular, there are annihilation
diagrams leading to the same final state (\ref{2HX}) in which a pair
$(Z^*,h^*)$ -- consisting of a virtual gauge boson and a virtual
Higgs boson -- is produced and, subsequently, $Z^*$ decays into
lepton pairs and $h^*$ radiates a real Higgs boson. Although none of
these topologies gives a dominant effect in front of the primary
gauge boson fusion mechanisms, a careful treatment is nevertheless
required in this case so as to preserve the consistency of the
overall procedure. In particular, the aforementioned Higgs strahlung
process demands the introduction of a Breit-Wigner propagator for
the virtual bosons.

The above considerations suggest that large production
cross-sections for the inclusive 2HX processes (\ref{2HX}) should be
possible at high energy regimes when both the exclusive 2H and 3H
channels (\ref{2H}) and (\ref{3H}) -- all of them mediated by
$Z$-boson exchange -- are kinematically suppressed as $\sim 1/S$. We
have undertaken this calculation using the computational tool
\emph{CompHEP} \cite{comphep}. In some steps of the computation, we
have also made use of \emph{FeynArts} and \emph{FormCalc}
\cite{feynarts}, which served also for cross-checking purposes.
Furthermore, to ease the comparison with the existing analysis of
the 3H processes (\ref{3H}), we present our numerical calculation on
the basis of the same set of free parameters as in \cite{previous},
i.e.
\begin{equation}\label{freep}
  (M_{\hzero},M_{\Hzero},M_{\Azero},M_{\Hpm},\tan\alpha,\tan\beta)\,.
\end{equation}
This means that the general 2HDM Higgs potential is treated under
the assumption that $\lambda_5=\lambda_6$ (in order not to depart
too much from the the MSSM structure of the Higgs sector, see
\cite{previous} for details). For a realistic computation, one has
to include all known existing constraints. For example, there are
strong limitations on the parameters of a type-II 2HDM coming from
FCNC processes, mainly from the charged Higgs boson contributions to
$\bsg$, which would overshoot the allowed experimental range unless
$M_{\Hpm}\gtrsim 350\,GeV$ \,\cite{gamba}; and there are also the
radiative corrections to $\delta\rho$ from the 2HDM sector, which
must not exceed the limit $|\delta\rho_{2HDM}|\le 10^{-3}$. Finally,
there are of course the direct search limits from LEP and the
Tevatron \cite{pdg08}. All these constraints are integrated in our
codes and, thus, the entire numerical analysis is consistent with
these bounds, along with the general unitarity constraints which
apply to both type-I and type-II 2HDM. More details on these
constraints are discussed in \cite{previous}, to which we refer the
reader for additional information. Let us also note that, in order
to obtain more accurate results, a running value for the
electromagnetic coupling constant $\alpha(M_Z) = 1/127.9$ has been
used.

%%%%%%%%%%%%%%%%%%%%%%%%%%%%%%%%%%%%%%%%%%%%%%%%%%%%%%%%%%%%%%%
%\noindent \subsection{Numerical analysis and discussion}
%\label{sect:2HXnumeric}
%%%%%%%%%%%%%%%%%%%%%%%%%%%%%%%%%%%%%%%%%%%%%%%%%%%%%%%%%%%%%%%

\subsection{Non-resonant double Higgs boson production}

For the numerical analysis, we will consider five different sets
(I-V) of parameters of the general 2HDM Higgs sector, see
Table~\ref{tab:mass_sets}. This should suffice to illustrate the
enhancement possibilities of the inclusive 2HX cross-section
(\ref{2HXF}). Notice that sets I-II and V are compatible with the
type-II 2HDM (because the charged Higgs boson mass is sufficiently
heavy to satisfy the aforementioned constraints). These sets are
actually possible for type-I models too. On the other hand, the
relatively light Higgs boson mass sets III-IV will be used
(exclusively) for the less constrained type I models. Let us also
remark that in the case of sets I-III the resonant decay
$\PHiggsheavyzero\to \PHiggslightzero\,\PHiggslightzero$ is not
possible, although for Sets I and III the mass threshold for such
resonant mode is closer than for Set II. Finally, Sets IV and V
explore the possibility of having some resonant Higgs boson decays
and are mainly intended for type-I and type-II models, respectively.
(Set V is also compatible with type-I models, as remarked before,
but we mainly aim at type-II ones for that set). We note that in the
case of Set IV, actually all of the resonant decays
$\PHiggsheavyzero\to \PHiggslightzero\PHiggslightzero,\,
\PHiggspszero\PHiggspszero,\,\PHiggs^+\PHiggs^-$  are kinematically
allowed whereas for Set V only the first decay is available.

\begin{table}[tbh]
\begin{center}
\begin{tabular}{|c||c|c|c|c|c|}
\hline  2HDM
 & Set I & Set II & Set III & Set IV & Set V \\
\hline\hline
$M_{h^0}\,[{\rm GeV}]$ & 100 & 150 & 150 & 100 & 125\\
$M_{H^0}\,[{\rm GeV}]$ & 190 & 250 & 290 & 225 & 280\\
$M_{A^0}\,[{\rm GeV}]$ & 360 & 360 & 150 & 110 & 300\\
$M_{H^\pm}\,[{\rm GeV}]$ & 350 & 350 & 150 & 105 & 350\\
\hline
\end{tabular}
\end{center}
\caption{\footnotesize{Higgs boson mass parameters used to discuss
phenomenologically relevant scenarios for the enhanced  2HX
production cross-sections (see the text). Sets III and IV are
possible only for type I models}.} \label{tab:mass_sets}
\end{table}

\begin{figure}[t]
\begin{center}
\vspace{0.3cm}
 \begin{tabular}{cc}
 \includegraphics[scale=0.3]{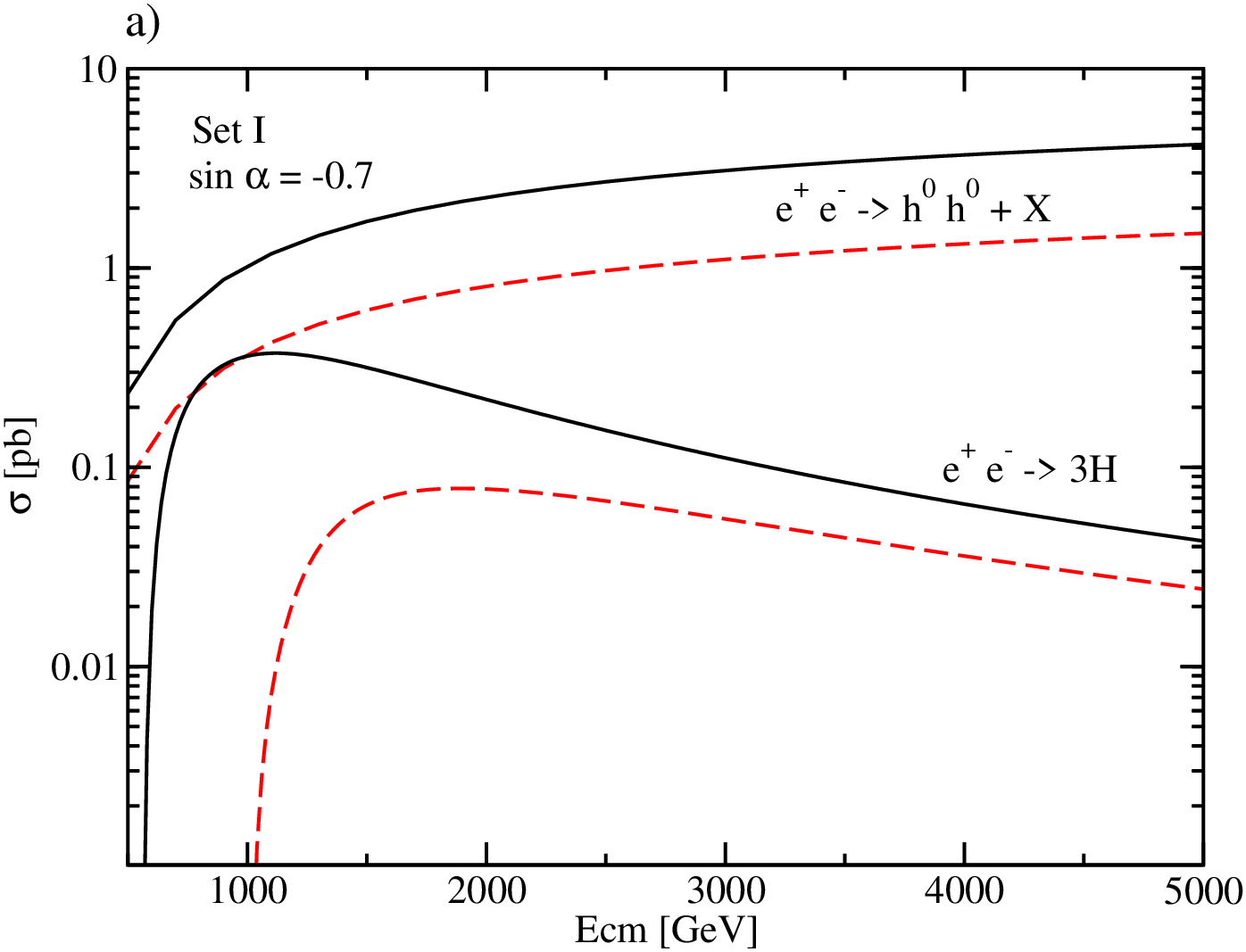} & \hspace{0.2cm}
\includegraphics[scale=0.3]{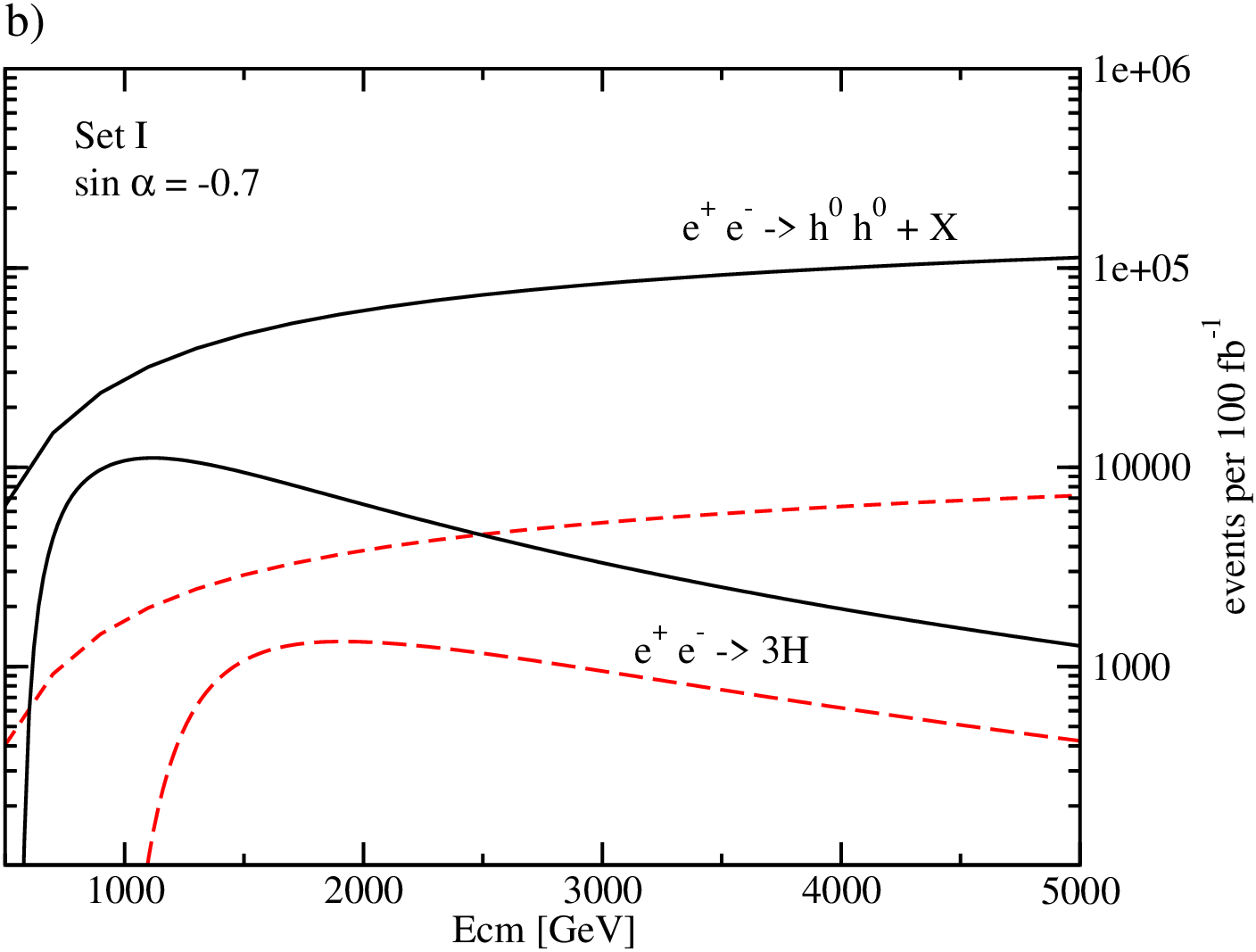}
 \end{tabular}
\end{center}
\caption{\footnotesize{\textbf{a)} Cross-section (in \picobarn) as a
function of $\sqrt{S}\equiv$ Ecm (in \GeV) for the particular 2HX
production process $\HepProcess{\APelectron \Pelectron \HepTo
\PHiggslightzero\PHiggslightzero+ X}$, together with the sum of all
the exclusive 3H channels, $\HepProcess{\APelectron \Pelectron
\HepTo 3\PHiggs}$ for a choice of Higgs masses as in Set I (in full
lines). We fix $\sin\alpha = -0.7$ and $\tan\beta = 25$, in which
case the relevant 3H couplings, and hence the production
cross-section, are maximally enhanced. The 2HX and 3H cross-sections
are also depicted (in dashed lines) as in the previous case but for
$M_{A^0}=M_{H^\pm}=800$~GeV; here, however, $\tan\beta \simeq 4$ in
order to preserve the consistency with the unitarity bounds. In the
right axis, we track the predicted number of events per $100 \invfb$
of integrated luminosity; \textbf{b)} As before, but for a lower
enhancement of the 3H couplings: $\tan\beta \simeq 14$ with Set I
masses (in full lines) and $\tan\beta \simeq 2$ with
$M_{A^0}=M_{H^\pm}=800$~GeV (in dashed lines).}} \label{fig:largema}
\end{figure}
%%%%%%%%%%%%%%%%%%%%%%%%%%%%%%%%%%%%%%%%%%%%%%%%%%%%%%%%%%%%%%%%%%%%%%
%%%%%%%%%%%%%%%%%%%%%%%%%%%%%%%%%%%%%%%%%%%%%%%%%%%%%%%%%%%%%%%%%%%%%%
\begin{figure}[t]
\begin{center}
 \begin{tabular}{ccc}
\includegraphics[scale=0.3]{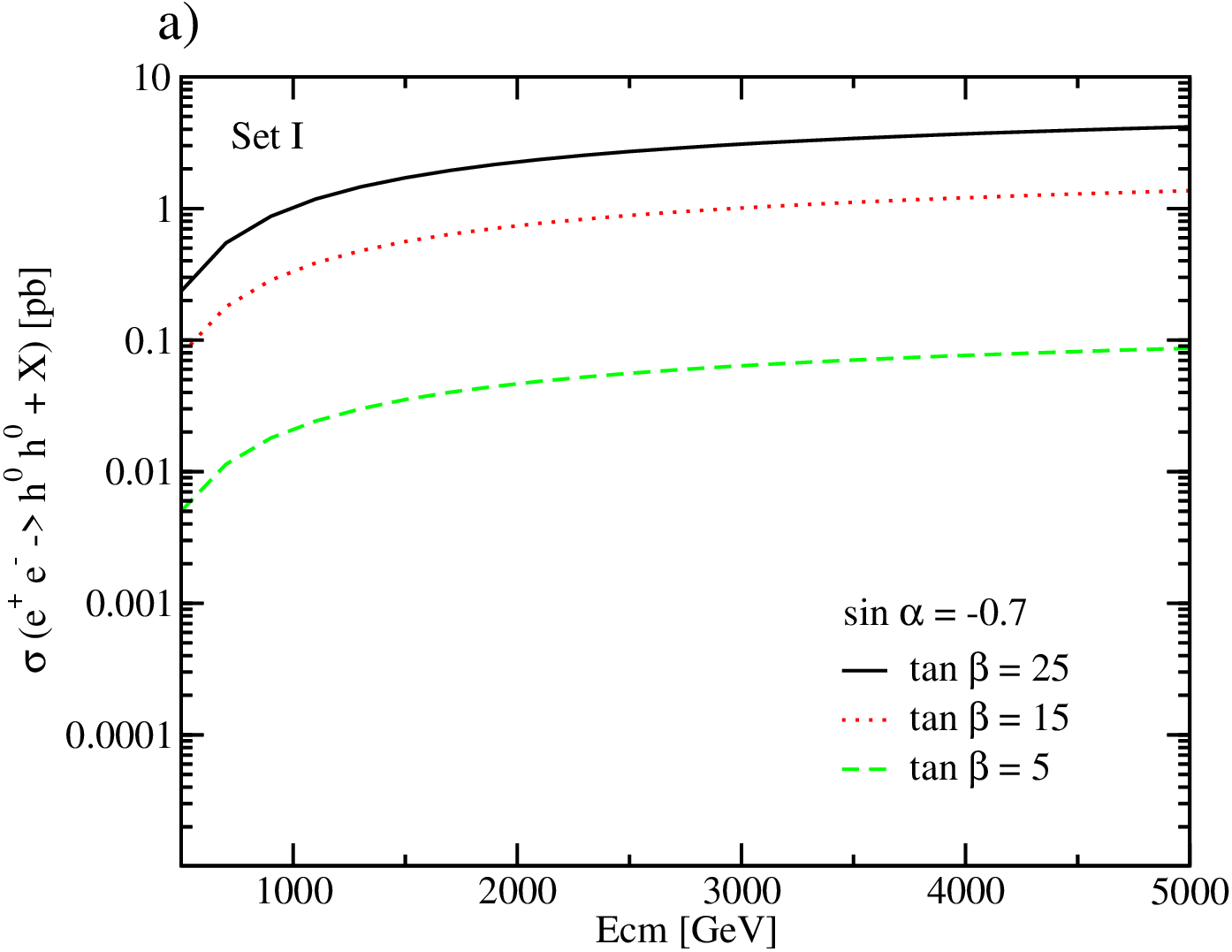}   & \hspace{0.2cm} &
\includegraphics[scale=0.3]{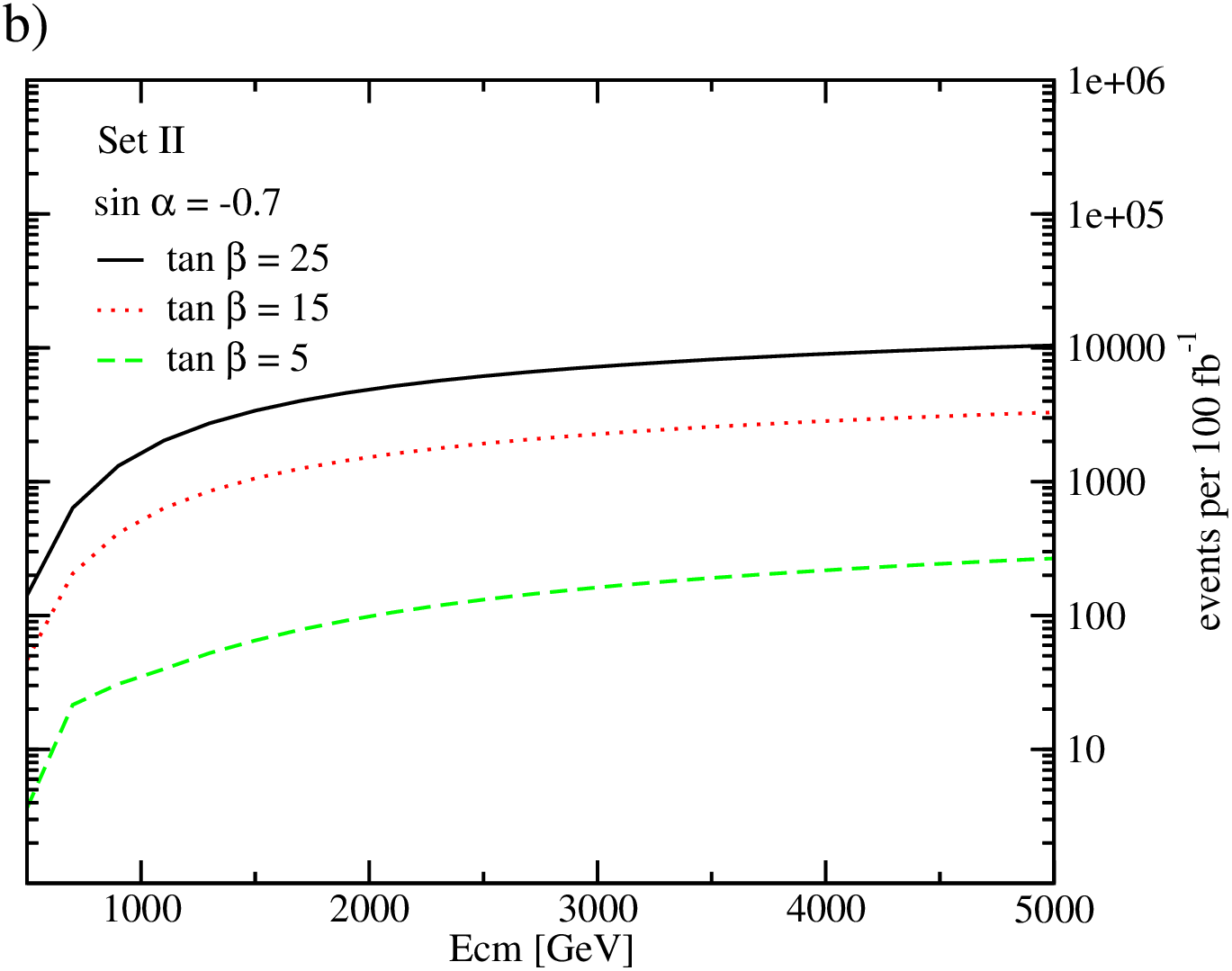}  \\ \, & \, \\
%\\ \, & \, \\
  \includegraphics[scale=0.3]{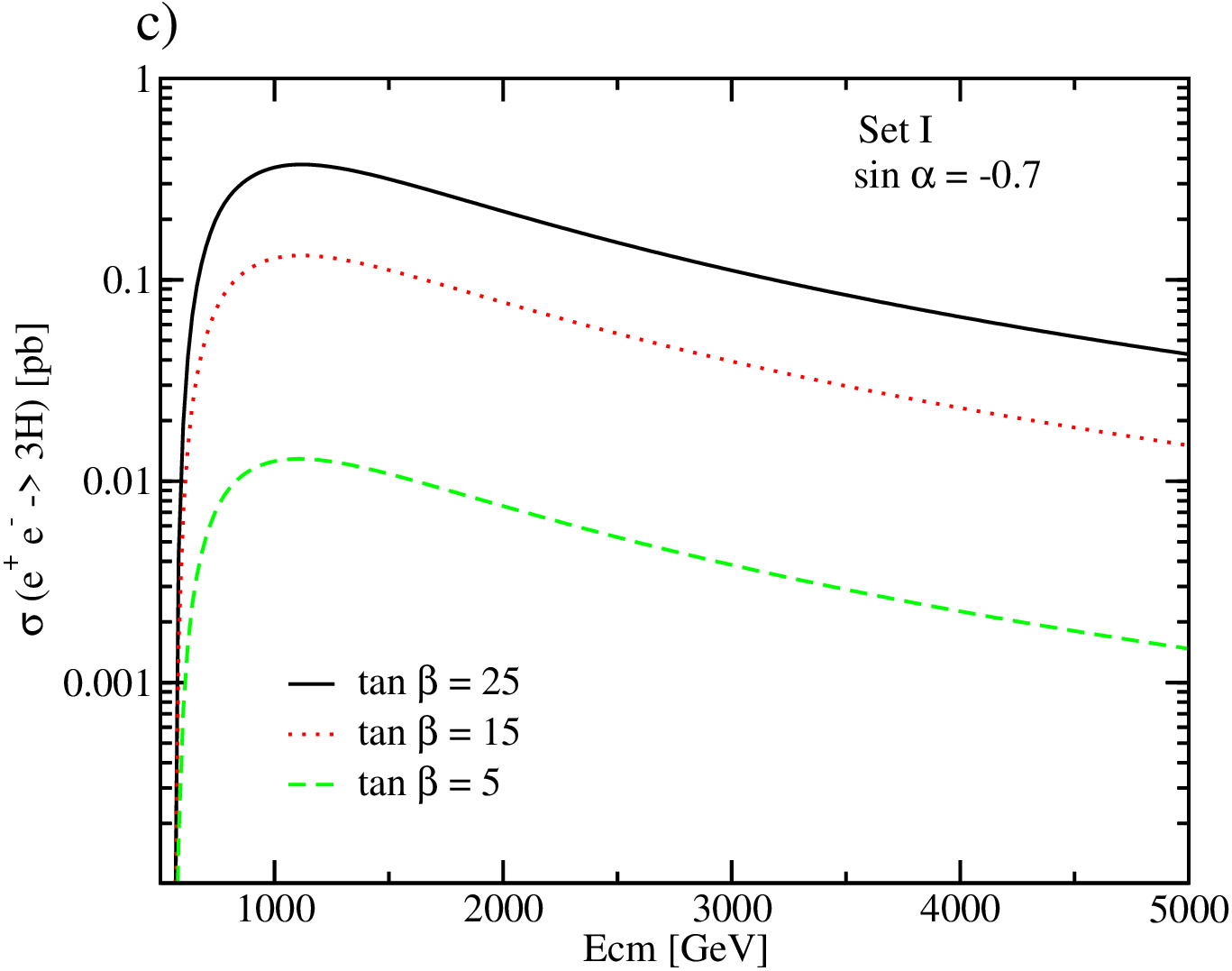}  & \hspace{0.2cm} &
\includegraphics[scale=0.3]{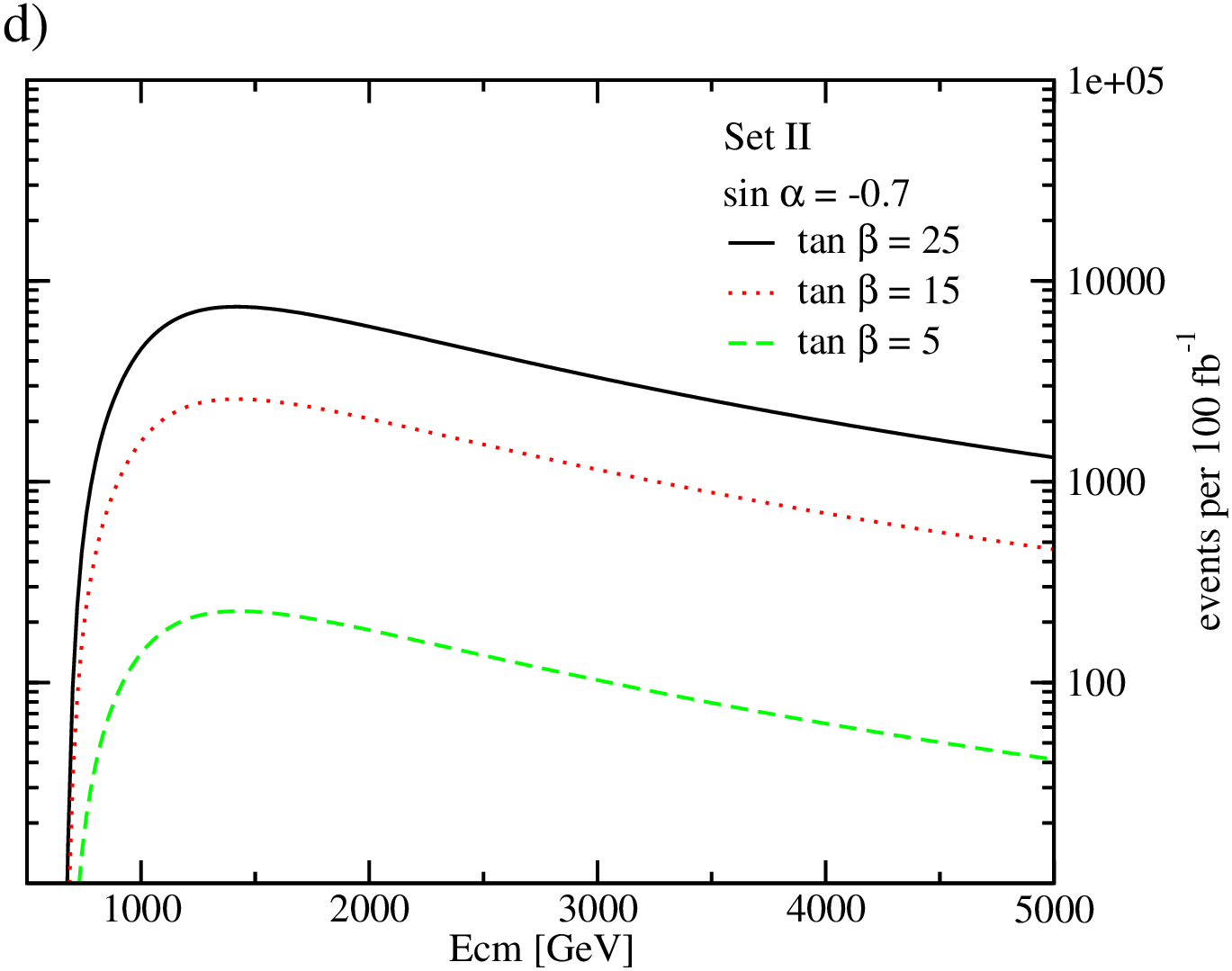}
 \end{tabular}
\end{center}
\caption{\footnotesize{Cross-section (in \picobarn) as a function of
$\sqrt{S}$ (in \GeV) for the inclusive Higgs boson-pair production
process
 $\HepProcess{\APelectron \Pelectron
\HepTo \PHiggslightzero\PHiggslightzero+ X}$ (upper panels) and for
the sum of all the exclusive 3H channels, $\HepProcess{\APelectron
\Pelectron \HepTo 3\PHiggs}$ (lower panels). In Figures a) and c),
Set I of Higgs boson masses is employed, while Set II is used for b)
and d) -- See Table~\ref{tab:mass_sets}. As in the previous figure,
we take $\sin\alpha = -0.7$ and we study the behavior of the
cross-section over different values of $\tan\beta$. In the right
axis of each plot, we track the predicted number of events per $100
\invfb$ of integrated luminosity.}} \label{fig:nonres}
\end{figure}
We consider first the numerical results obtained from the
non-resonant scenarios, which are more general. In doing so, we wish
to compare the cross-sections for both the 3H and 2HX channels
(\ref{3H}) and (\ref{2HX}), respectively.  In
Fig.~\ref{fig:largema}a, we show these production cross-sections as
a function of the CM energy for Set I of Higgs boson masses (full
lines in that figure).  The fixed value of the $\CP$-even mixing
angle in this figure ($\sin\alpha = -0.7$) has been determined in
combination with $\tan\beta$ after maximizing the cross-sections for
the mass Set I under the various constraints discussed previously.
The relevant trilinear Higgs self-couplings
$\PHiggslightzero\PHiggslightzero\PHiggslightzero$ and
$\PHiggsheavyzero\PHiggslightzero\PHiggslightzero$, and hence the
overall production cross-section, become maximally enhanced at
$\tan\beta\simeq 25$. The main constraint that fixes the
aforementioned $\tan\beta$ value is the unitarity bound of the
trilinear Higgs couplings. In order to explore the effect on the
cross-sections after significantly increasing some masses (while
respecting all the necessary bounds mentioned above), we also show
the corresponding 2HX and 3H cross-sections for
$M_{A^0}=M_{H^\pm}=800$~GeV, with all other parameters fixed as in
Set I (dashed curves in that figure). In this case, the unitarity
constraints pull the maximum value of $\tan\beta$ down to $\sim 4$.
{Noteworthy in Fig.~\ref{fig:largema} is that the kinematical
threshold for the overall 3H contribution shifts to higher energies
when we boost $M_{\PHiggspszero}$ up to 800 \GeV. Let us clarify
that the rise of further thresholds does not leave a visible
footprint on $\sigma(\sqrt{S})$, provided one of the 3H channels (in
the present case $\PHiggslightzero\PHiggslightzero\PHiggspszero$)
dominates over the remaining ones. The subdominant channels,
although have less enhanced 3H couplings and a larger phase-space
suppression, contribute to smooth out significantly the damping of
the total 3H cross-section as a function of $\sqrt{S}$ as compared
to the individual channels.}

In Fig.~\ref{fig:nonres}, we show the numerical analysis of the 2HX
cross-section for various values of $\tan\beta$, with the Higgs mass
parameters taken as in Set I (Fig.~\ref{fig:nonres}a) and Set II
(Fig.~\ref{fig:nonres}b). The corresponding 3H production
cross-sections are shown for comparison in the lower panels
(Figs.~\ref{fig:nonres}c and \ref{fig:nonres}d). Likewise, in
Fig.~\ref{fig:largema}b we display the corresponding results
obtained for a lower enhancement of the 3H couplings (viz. up to one
half of the standard unitarity bound used in\,\cite{previous}).
Notice that our energy scan actually sweeps a wide range which
reaches up to $5$ TeV in order to better show the asymptotic trend
of the cross-sections. In practice, the ILC will cover only the
approximate range $0.5-1.5$ TeV, whereas CLIC may reach
$3$~TeV\,\cite{ILC-CLIC}.

The dominant effect on the inclusive 2HX amplitudes is proportional
to the Higgs trilinear couplings
$\PHiggsheavyzero\PHiggslightzero\PHiggslightzero$ and
$\PHiggslightzero\PHiggslightzero\PHiggslightzero$.  These couplings
are enhanced at large values of $\tan\beta$, {as can be seen from
the first Eq.\,(\ref{eq:3Hcoup})} and in general from Table 1 of
Ref.\,\cite{previous}.
{Let us also emphasize that, in order to } better appreciate the
$\tan\beta$-dependence, the maximum 2HX and 3H cases corresponding
to $\tan\beta=25$ are again included in Figs.~\ref{fig:nonres}a,c
along with the other (smaller) $\tan\beta$ values. Although the
overall production rates decrease with decreasing $\tan\beta$, the
2HX channel remains dominant at (and above) $\sqrt{S}=1$~TeV for Set
I. In this same energy range, but for Higgs boson masses as in Set
II, the 2HX channel remains comparable to the 3H channel only for
the largest allowed values of $\tan\beta$. At higher energies,
however, such as those planned for CLIC, the 2HX channels become the
leading ones even at small values of $\tan\beta$.
{To be sure, some of these gauge fusion processes furnish a very
competitive strategy to probe the 3H self-couplings. This strategy
nicely complements the prospects that were singled out for the
exclusive 3H channels in Ref.~\cite{previous}, particularly at very
large center-of-mass energies, where the 3H signal is greatly
suppressed whilst the 2HX one remains sustained and even
logarithmically enhanced.}
Cross-sections for the remaining 2HX final states
containing heavier Higgs bosons are not shown in
Fig.~\ref{fig:nonres} as they are found to have negligible
production rates when compared to the corresponding 3H final states
in this scenario.  Explicitly, the maximum production rates are
found to be of the order $10^{-2}$~\picobarn\ for the
$\PHiggslightzero\PHiggsheavyzero$ case, and below
$10^{-3}$~\picobarn\ for the rest of the 2HX final states.

\begin{figure}[t]
\begin{center}
 \begin{tabular}{ccc}
 \includegraphics[scale=0.3]{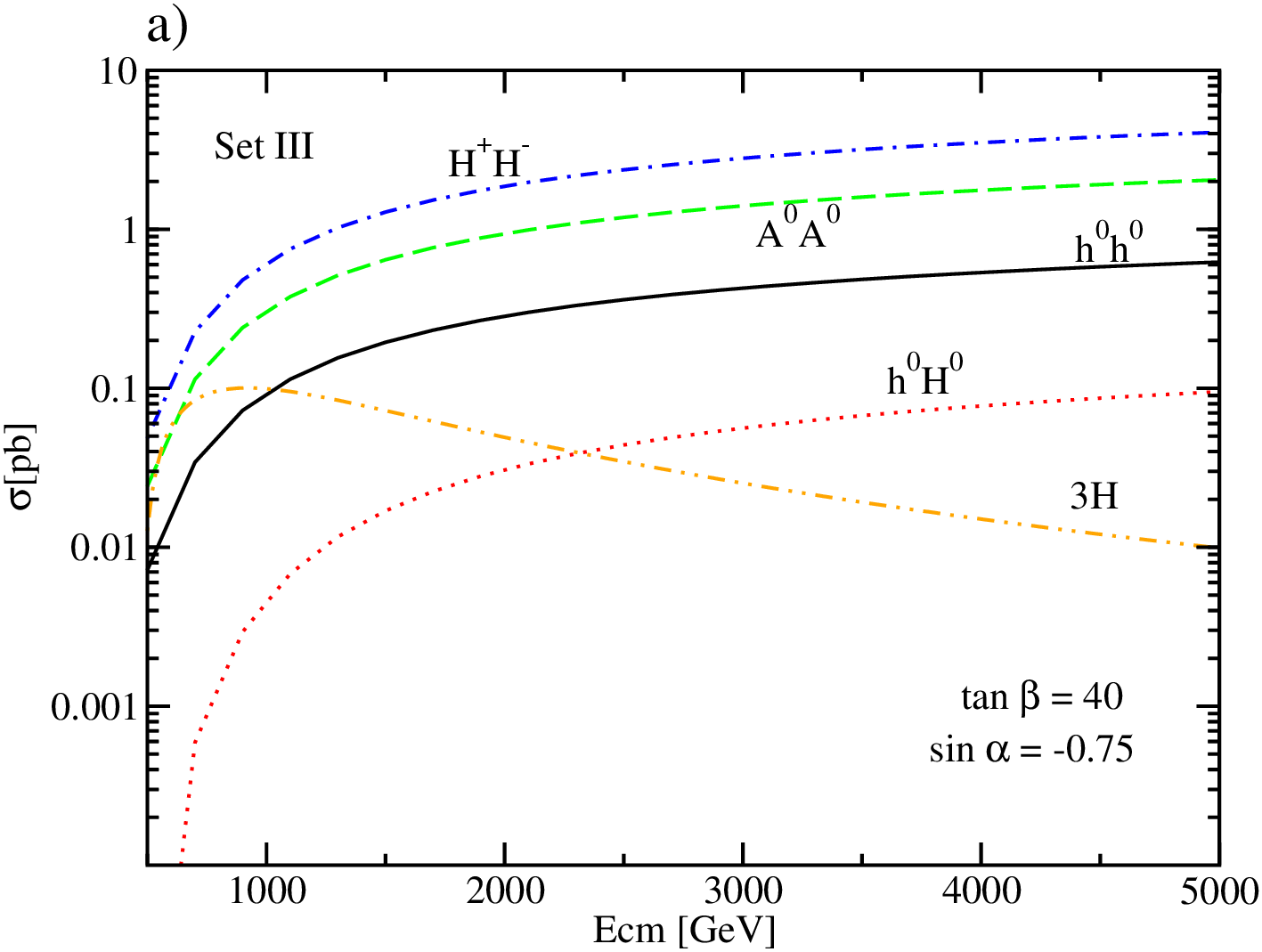} & \hspace{0.2cm} &
\includegraphics[scale=0.3]{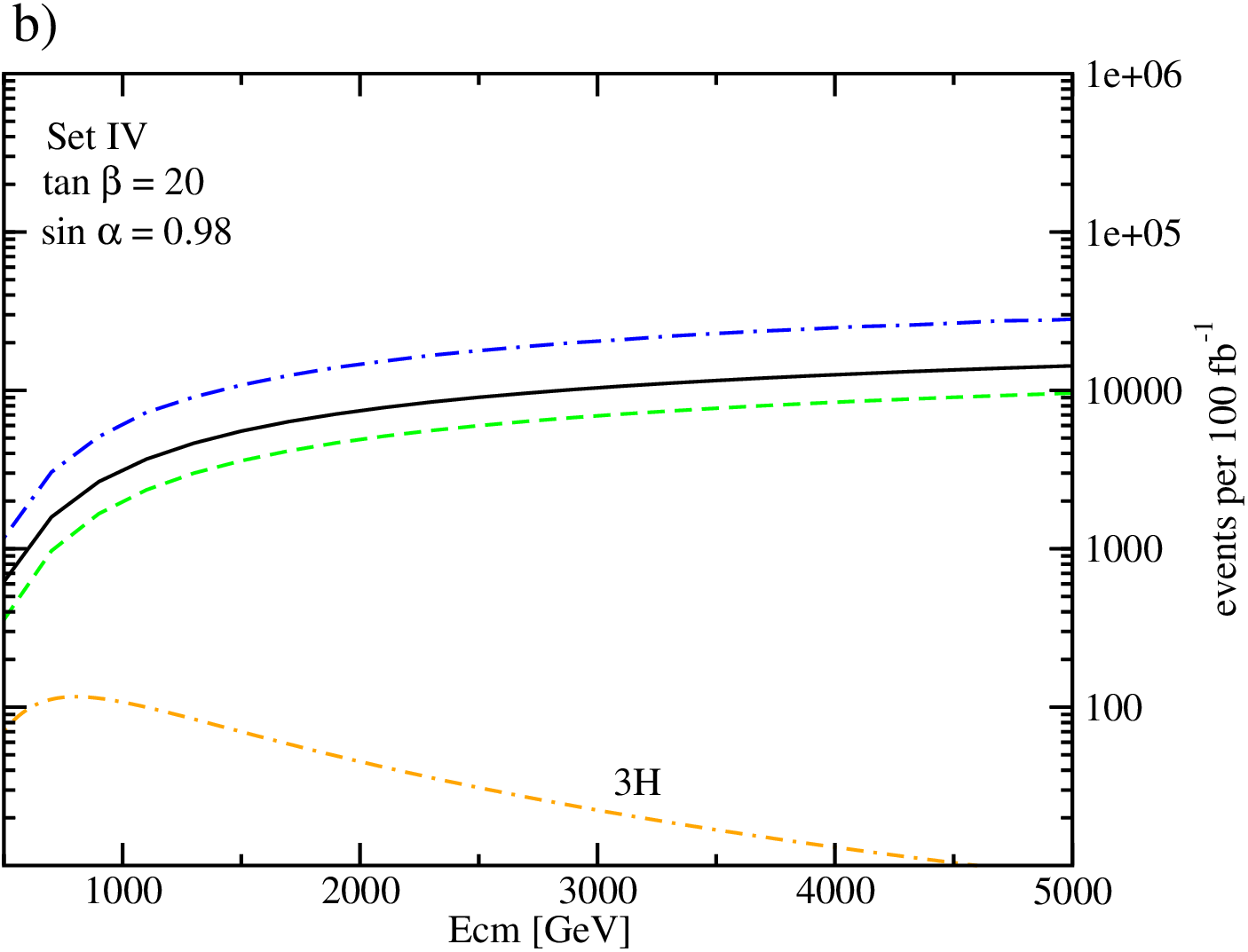}
 \end{tabular}
\end{center}
\caption{\footnotesize{Cross-section (in \picobarn) as a function of
$\sqrt{S}$ (in \GeV) for the 2HX processes $\HepProcess{\APelectron
\Pelectron \HepTo hh + X}$ for Sets III and IV of Higgs boson
masses, which are suitable for type-I 2HDM -- see
Table~\ref{tab:mass_sets}. Displayed are the Higgs boson-pair
channels whose cross-sections lie above $0.01 \,\picobarn$, together
with the corresponding triple-Higgs production rate,
$\APelectron\Pelectron\to 3H$, for each scenario. The values chosen
for $\sin\alpha$ and $\tan\beta$ are quoted in the figures.}}
% in the non-resonant case (upper panels) and for
% the sum of all the exclusive 3H channels, $\HepProcess{\APelectron
% \Pelectron \HepTo 3\PHiggs}$ (lower panels). Pictures a) and c)
% present the results for different values of $\sin\alpha$ at fixed
% $\tan\beta$, and conversely in panels b), d). In the right axis we
% track the predicted number of events per $100 \invfb$ of integrated
% luminosity. Set IV of Higgs boson masses has been used throughout, cf.
% Table~\ref{tab:mass_sets}.}
\label{fig:type1}
\end{figure}

Notwithstanding, a starkly distinct panorama shows up for lighter
$\PHiggs^\pm$ and $\PHiggspszero$ bosons. In
Figure~\ref{fig:type1}a, we display $\sigma(\sqrt{S})$ corresponding
to Set III of Higgs boson masses (see Table~\ref{tab:mass_sets}), in
which case sizable cross-sections are attained for a number of 2HX
channels: $\PHiggs^+\PHiggs^-$ yields $\,\sim 1$\,\picobarn;
$\PHiggslightzero\PHiggslightzero$ and $\PHiggspszero\PHiggspszero$
give $\,\sim 0.1$~\picobarn, and $\PHiggslightzero \PHiggsheavyzero$
renders $\,\sim 0.01$~\picobarn. Owing to the relatively light mass
of the charged Higgs boson, the latter scenario is only allowed for
type-I, not for type-II, 2HDM -- otherwise it would yield an
exceedingly large FCNC contribution to $\bsg$ \cite{pdg08}.
Similarly, in Fig.~~\ref{fig:type1}b we present the corresponding
results for Set IV of Higgs boson masses, which also applies
exclusively for type-I models, but refers to the resonant situation
in which the on-shell decays $\PHiggsheavyzero\to
\PHiggslightzero\PHiggslightzero,\,
\PHiggspszero\PHiggspszero,\,\PHiggs^+\PHiggs^-$ are all allowed. In
the next section, we further dwell on the resonant case, but we
consider a scenario (valid for both type-I and type-II models) where
only the final state with two light CP-even Higgs bosons is
permitted, i.e. $\PHiggsheavyzero\to
\PHiggslightzero\PHiggslightzero$, and we study it in more detail.

%%%%%%%%%%%%%%%%%%%%%%%%%%%%%%%%%%%%%%%%%%%%%%%%%%%%%%%%%%%%%%%%%%%%%%%%%
%%%%%%%%%%%%%%%%%%%%%%%%%%%%%%%%%%%%%%%%%%%%%%%%%%%%%%%%%%%%%%%%%%%%%%%%%
\begin{figure}[t]
\begin{center}
 \begin{tabular}{ccc}
 \includegraphics[scale=0.3]{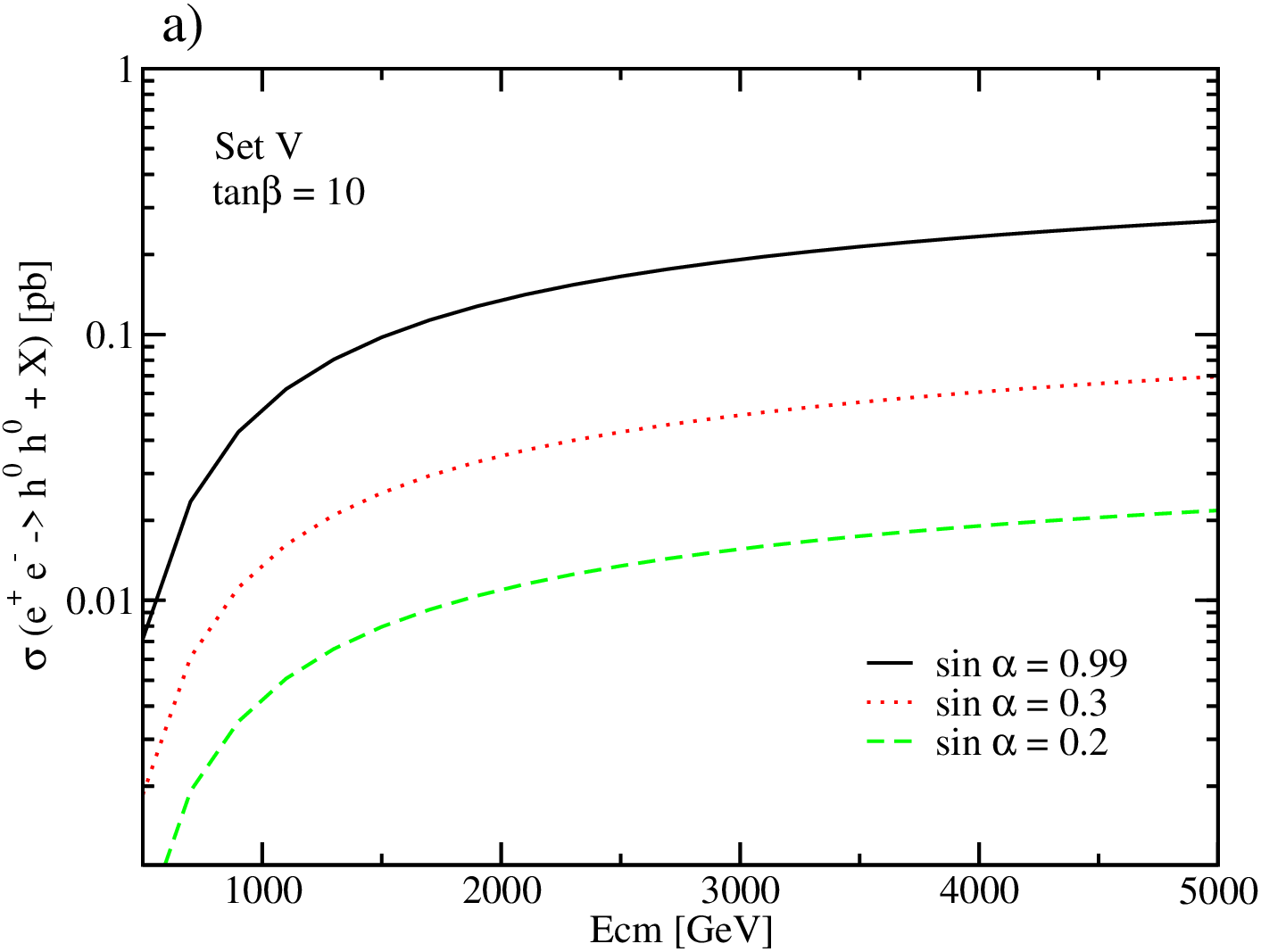} & \hspace{0.2cm} &
\includegraphics[scale=0.3]{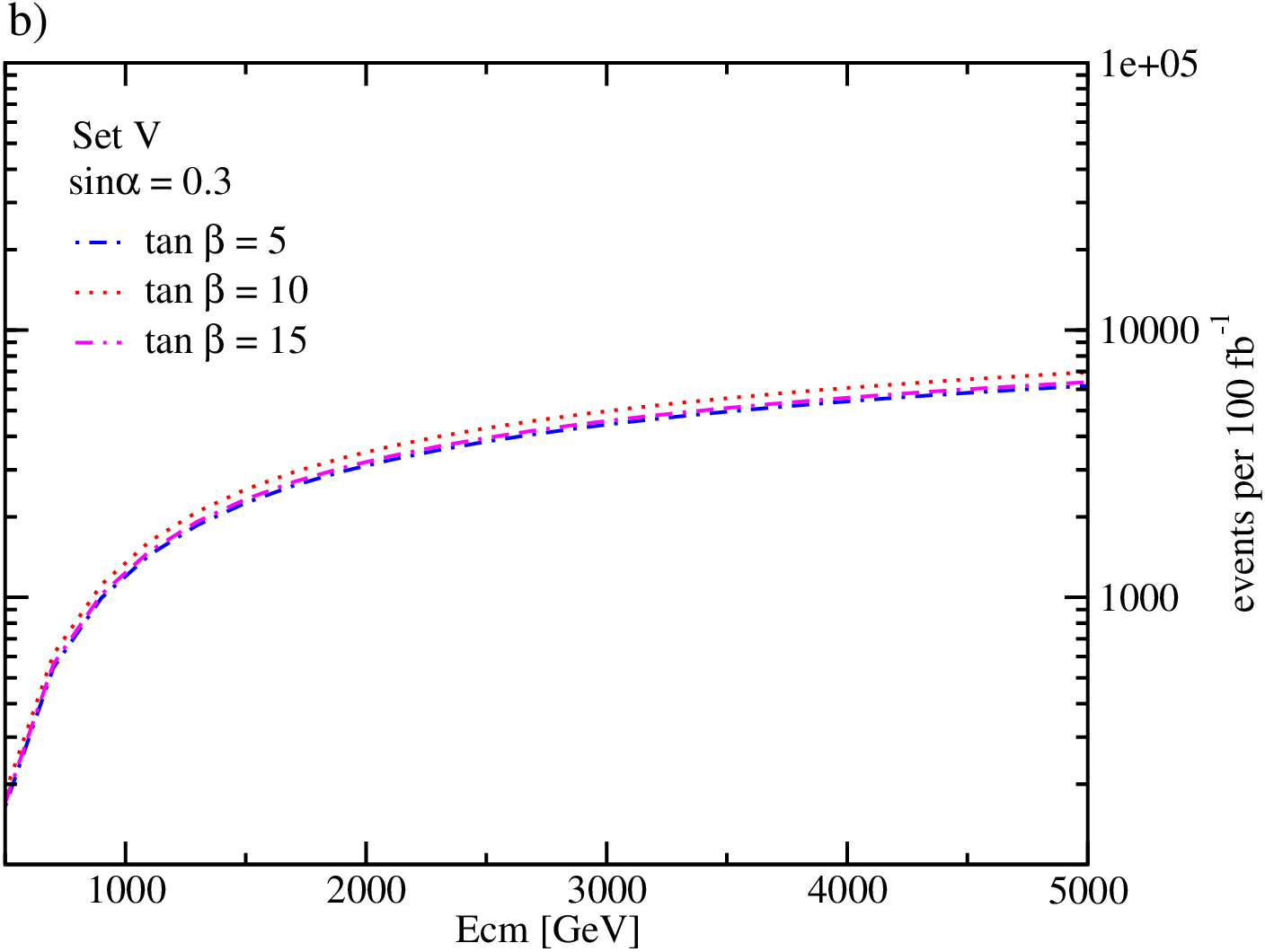}  \\ \, & \, \\
%\\ \, & \, \\
  \includegraphics[scale=0.3]{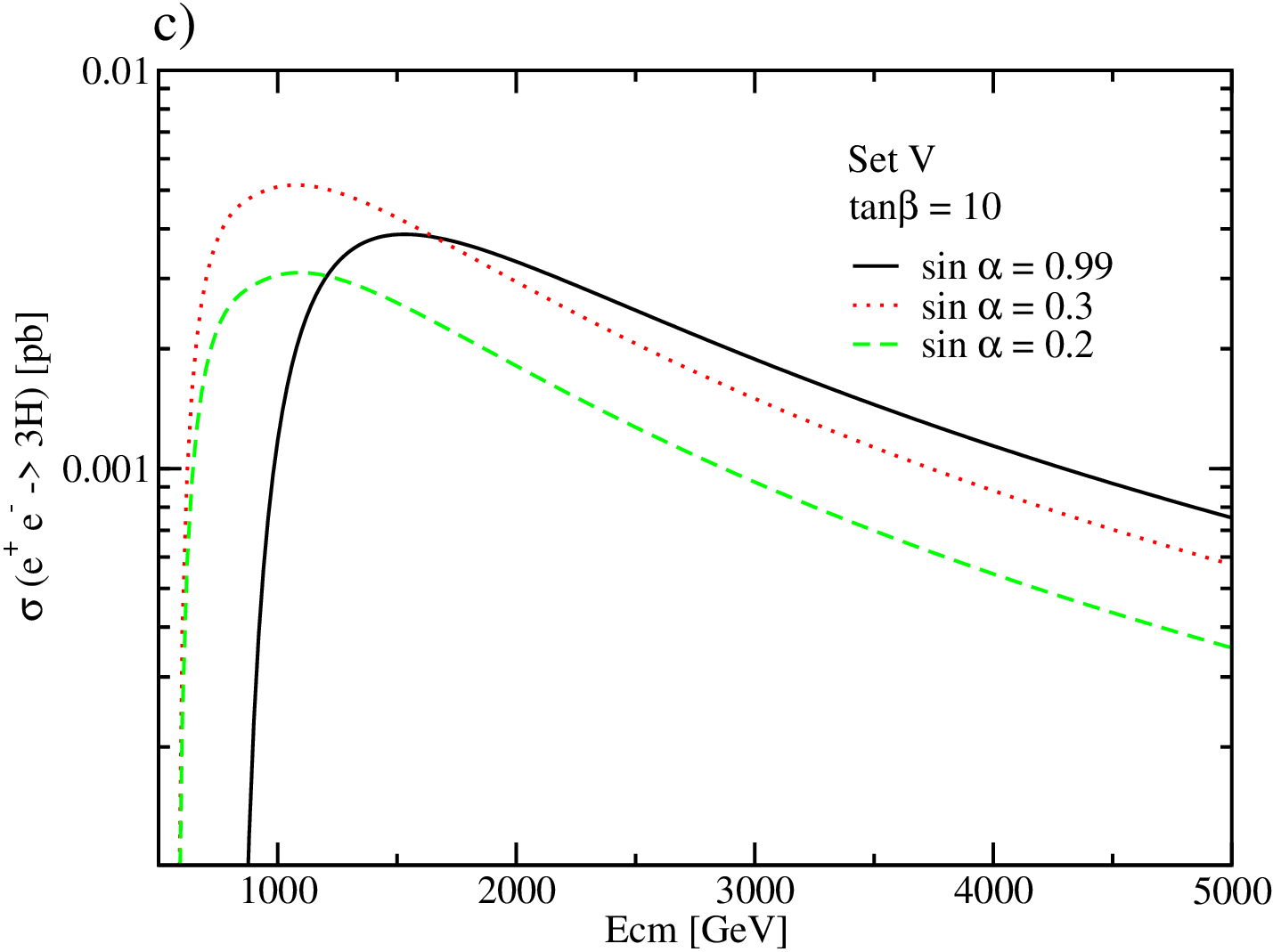} & \hspace{0.2cm}
& \includegraphics[scale=0.3]{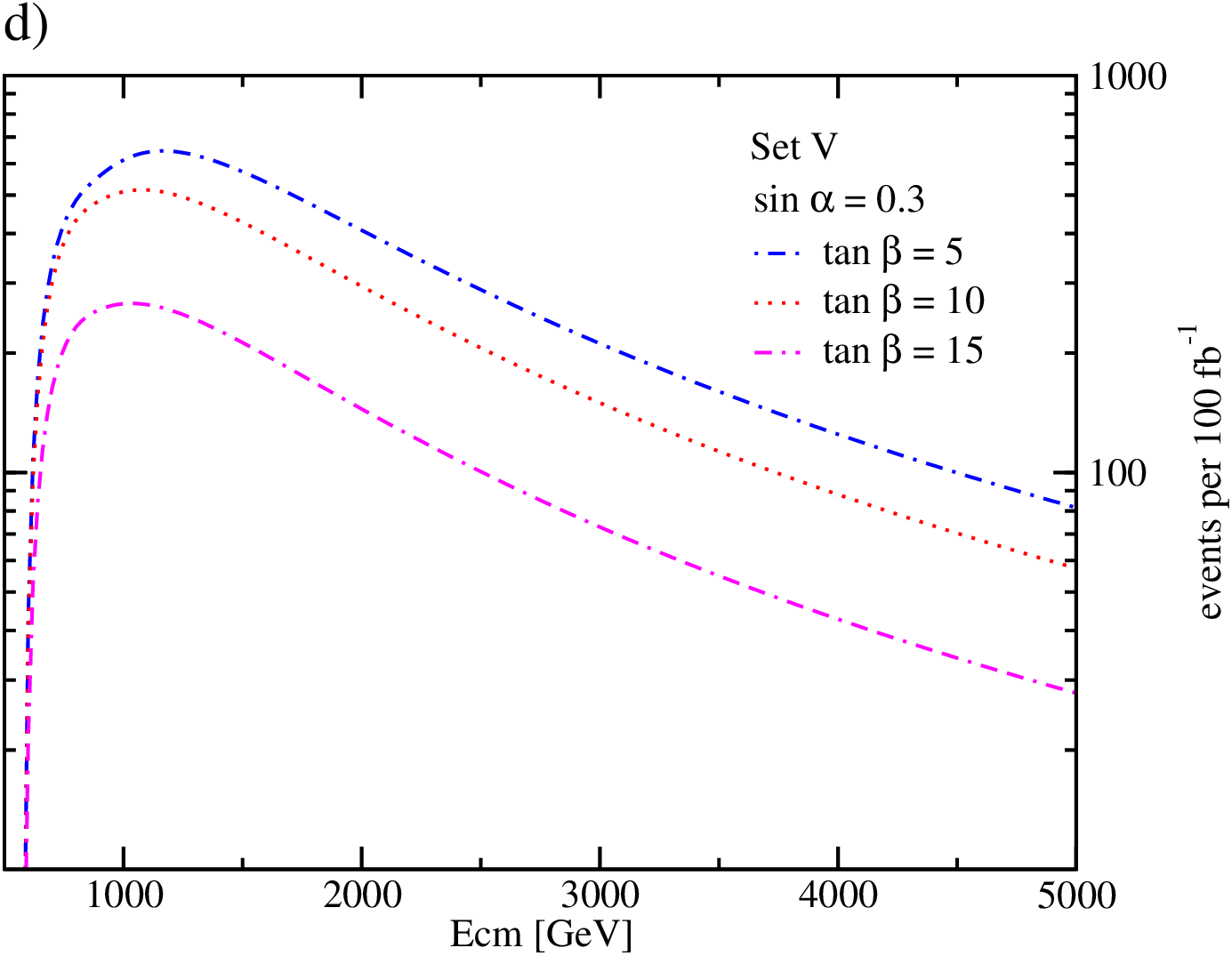}
 \end{tabular}
\end{center}
\caption{\footnotesize{Cross-section (in \picobarn) as a function of
$\sqrt{S}$ (in \GeV) for $\HepProcess{\APelectron \Pelectron \HepTo
\PHiggslightzero \PHiggslightzero + X}$ in the resonant case (upper
panels) and for the sum of all the 3H channels
$\HepProcess{\APelectron \Pelectron \HepTo 3\PHiggs}$ (lower
panels). Pictures a) and c) present the results for different values
of $\sin\alpha$ at fixed $\tan\beta$, and conversely in panels b),
d). In the right axis, we track the predicted number of events per
$100 \invfb$ of integrated luminosity. Set V of Higgs boson masses
has been used throughout.}} \label{fig:res}
\end{figure}

%\pagebreak
\mysubsection{Resonant double Higgs boson production}

We note that the scenarios considered above are proper of the
general 2HDM. In the MSSM case, where the masses of the $\CP$-even
light and heavy Higgs bosons $\PHiggslightzero,\PHiggsheavyzero$ are
not independent parameters once $\tan\beta$ and $M_{\Azero}$ are
given\,\cite{hunter}, the mass splittings indicated in
Table~\ref{tab:mass_sets} are not possible. For larger enough values
of $M_{\PHiggsheavyzero}$, there is a drastic change in the
behaviour for the production cross-section of the inclusive channel
$\HepProcess{\APelectron \Pelectron \HepTo
\PHiggslightzero\PHiggslightzero + X}$ since the on-shell decay
$\PHiggsheavyzero\to \PHiggslightzero\PHiggslightzero$ becomes
kinematically available. Indeed, with
$M_{\PHiggsheavyzero}>2M_{\PHiggslightzero}$ the cross-sections are
somewhat less dependent upon the enhancement of the Higgs trilinear
couplings, the reason being that highly enhanced trilinears now lead
to a dramatic broadening of the $\PHiggsheavyzero$ resonance with a
branching ratio essentially of order one.

In Figures~\ref{fig:res}a-\ref{fig:res}d, we exhibit a panoply of
2HX and 3H production cross-sections for the Higgs boson masses as
in Set V of Table~\ref{tab:mass_sets}. We explore the dependence
with the CM energy and the mixing angles of the Higgs sector. In all
cases, it corresponds to a situation where the on-shell decay
$\PHiggsheavyzero\to \PHiggslightzero\PHiggslightzero$ is possible
but, in contradistinction to Set IV considered in
Fig.~\ref{fig:type1}b, the alternate decays $\PHiggsheavyzero\to
\PHiggspszero\PHiggspszero,\,\PHiggs^+\PHiggs^-$ are not allowed.
The case under consideration is more along the line of type-II
models, which are the the closest ones to the SUSY case. From these
figures, it is patent that the inclusive cross-section exceeds the
3H channel for all energies above the TeV scale. Furthermore, there
is a significant dependence of the 2HX cross-section on
$\sin\alpha$, which enters through the vertex factor
$\cos^2(\beta-\alpha)$ associated to the on-shell production of
$H^0$ from $W^+W^-$ and $Z^0Z^0$ fusion (cf.
Fig.\,\ref{Feynman_diagram}). At the same time, the 2HX production
cross-sections are now largely independent of $\tan\beta$, because
changing this parameter just leads to small fluctuations of the
$H^0$ branching ratio around one. At variance with this mild
$\tan\beta$ dependence of the 2HX processes, the main enhancement
source of the 3H final states still resides in the $\tan\beta$
effective dependence of the Higgs trilinears, as can be seen in
Figure~\ref{fig:res}d.

%\jump
Interestingly enough, let us emphasize that the potentially large
values of the 2HX cross-sections studied up to now, either with
resonant or non-resonant configurations, are a kind of trademark
prediction of the \emph{non}-supersymmetric two-Higgs-doublet
models. Of course, the collection of diagrams shown in
Fig.~\ref{Feynman_diagram} also accounts for the corresponding 2HX
processes within the MSSM. Nonetheless, the 3H couplings are
constrained by supersymmetry and are directly related to the
electroweak gauge couplings; there is therefore no possibility of
enhancement. Furthermore, in the MSSM the region of parameter space
where the relation $M_{\PHiggsheavyzero}>2\,M_{\PHiggslightzero}$
holds is dominant. As a consequence, the inclusive 2HX production
will be brought about predominantly by the production of on-shell
$\PHiggsheavyzero$ bosons via $\PW^+\PW^-$ fusion (cf.
Fig.\,\ref{Feynman_diagram}) and followed by their subsequent
on-shell decay
$\PHiggsheavyzero\to\PHiggslightzero\PHiggslightzero$. As already
mentioned, it is precisely the $\PW^+\PW^-\PHiggsheavyzero\sim
\cos(\beta-\alpha)$ coupling that limits the resonant 2HX production
in the MSSM because supersymmetry constrains the lightest $\CP$-even
Higgs boson $\PHiggslightzero$ to have SM-like couplings when the
remaining partners are heavy. Consequently, the resulting
cross-sections are expected to lie far below the {optimal} 2HDM
yields. To illustrate these features in a concrete way, in
Table~\ref{tab:mssm} we compute the predicted MSSM cross-sections
for the inclusive production of a $\PHiggslightzero\PHiggslightzero$
pair, $\sigma(\APelectron\Pelectron
\to\PHiggslightzero\PHiggslightzero + X)$. {Concerning the CM
energies, we assume the operation range that is scheduled for either
the ILC ($\sqrt{S}= 0.5 - 1.5\,~\TeV$) and CLIC (up to
$\sqrt{S}=3$~TeV)}, and take three different $\CP$-odd Higgs boson
masses $M_{\PHiggspszero} = 200\, , \,300\,$ and $500$~GeV. The
value of $\tan\beta$ is not shown, but it is determined such that to
(approximately) optimize the corresponding production rate. In
calculating these values, we have taken all soft SUSY-breaking
masses equal to $M_{\rm SUSY}=1$~TeV, along with $\mu=200$~GeV and
$A_t=A_b=A_\tau=1$~TeV
%
%, while
%keeping the remaining MSSM parameter at the values displayed
%in Table~\ref{tab:mssmpar}
\footnote{For the determination of the MSSM Higgs sector and the
mixing angle $\alpha$ we make use of the \emph{FeynHiggsFast} code,
which is included by default in the \emph{CompHEP} setup
\cite{feynhiggsfast}.}. From Table~\ref{tab:mssm}, we see that the
2HX rates render a contribution of order of a few \femtobarn\,, at
most, for $M_{\PHiggspszero} > 300\,\GeV$. Nevertheless, there is a
narrow corner in the region of low $M_{\PHiggspszero}$/low
$\tan\beta$ (which is severely restricted by the LEP mass bounds
\cite{pdg08}) wherein the $\PW^+\PW^-\PHiggsheavyzero$ coupling is
not so much hampered and hence the cross-section climbs up to $\sim
\,10\,\femtobarn$. {By comparing the (optimal) values in
Table~\ref{tab:mssm} to the most favorable scenarios displayed in
{Figures \ref{fig:largema}-\ref{fig:res}}, we conclude that the 2HX
signal in the 2HDM is typically a factor $10-100$ larger than its
MSSM counterpart.}

\begin{table}[t]
\begin{center}
\begin{tabular}{|c||c|c|c|}
\hline \, & $M_{\PHiggspszero} = 200\,\GeV$ & $M_{\PHiggspszero} =
300\,\GeV$ & $M_{\PHiggspszero} = 500\,\GeV$ \\ \hline
% sqrtS = 0.5 TeV
$\sigma(\sqrt{S}= 0.5 \,\TeV) (\picobarn)$ & $1.5\times10^{-3}$ &
$9.0\times10^{-5}$ & $4.2\times10^{-5}$ \\ \hline
% sqrtS = 1 TeV
$\sigma(\sqrt{S}= 1.0 \,\TeV) (\picobarn)$ & $3.3\times10^{-3}$ &
$5.6\times10^{-4}$ & $2.0\times10^{-4}$ \\ \hline
% sqrtS = 1.5 TeV
$\sigma(\sqrt{S}= 1.5 \,\TeV) (\picobarn)$ & $5.7\times10^{-3}$ &
$1.1\times10^{-3}$ & $4.7\times10^{-4}$ \\ \hline
% sqrtS = 3 TeV
$\sigma(\sqrt{S}=
3.0 \,\TeV) (\picobarn)$ & $1.1\times10^{-2}$ & $2.7\times 10^{-3}$ & $1.5\times10^{-3}$\\
\hline
\end{tabular}
\end{center}
\caption{\footnotesize{Maximum cross-section
$\sigma(\APelectron\Pelectron \to\PHiggslightzero\PHiggslightzero +
X)$ in various scenarios within the MSSM. In all of them we find
that the optimal cross-section value corresponds to
$\tan\beta\gtrsim 2$.}} \label{tab:mssm}
\end{table}

%\begin{table}[t]
%\begin{center}
%\begin{tabular}{|c|c|}
%\hline $M_{SUSY}\ {\GeV}$ & 1000  \\ \hline $\mu\ {\GeV}$ & 200  \\
%\hline $A_t\ {\GeV}$ & 1000  \\ \hline $A_b \ {\GeV}$ & 1000  \\
%\hline $A_\tau \ {\GeV}$ & 1000  \\ \hline
%\end{tabular}
%\end{center}
%\caption{\footnotesize{Choice of SUSY parameters used for the computation of the
%Higgs boson masses of the inclusive 2H production within the MSSM.
%As in Ref.\cite{previous}}} \label{tab:mssmpar}
%\end{table}

Finally, we have also evaluated the ``SM background'', i.e. the
cross-section for double Higgs production from gauge fusion in the
SM for the same $\sim$TeV energies under study. We find e.g. that
$\sigma(\APelectron\Pelectron \to V^*V^*\to H\,H \,+ X)= 10^{-3} -
10^{-5}$ pb for SM Higgs boson masses in the range $M_H=115 -
300\,\GeV$ (see Table ~\ref{tab:sm}), which is relatively quite
small as compared to our favorite 2HDM scenarios. Furthermore, we
note that the produced SM Higgs boson will predominantly decay into
gauge boson pairs $W^+W^-$, $Z^0Z^0$ and (to a lesser extent) to
bottom quark pairs $b\bar{b}$ and top quark pairs $t\bar{t}$ (if
kinematically possible). Subsequently, the gauge bosons will decay
both into leptonic and light quark channels. In contrast, in the
2HDM case, the very same conditions that favor the fusion production
of Higgs bosons do also favor the decay of the produced Higgs boson
into other Higgs bosons and these into heavy quarks. So the kind of
signature is very distinct. Another source of background to the 2HDM
signal could come from gauge boson fusion into gauge bosons,
essentially $\PZ^0$ pairs. If these gauge bosons subsequently decay
into quarks, this would mimic the Higgs boson themselves decaying
into quark pairs. However, explicit calculation shows that the
cross-section $\sigma(\APelectron\Pelectron \to V^*V^*\to
\PZ^0\,\PZ^0 \,X)$ reaches up to $0.1$\picobarn\, at most (Cf. Table
~\ref{tab:sm}). Hence it lies one order of magnitude below the
favorite 2HX cases. For the less favorable situations, however,
further studies on the final jet distribution might be necessary to
disentangle the corresponding signatures.
\begin{table}[t]
\begin{center}
\begin{tabular}{|c||c|c||c|c|}
\hline Background & \multicolumn{2}{c||}{$M_{\PHiggs_{SM}} =
115\,\GeV$} & \multicolumn{2}{|c|}{$M_{\PHiggs_{SM}} = 300\,\GeV$}
\\\cline{2-5}
 & $\PHiggs_{SM}\,\PHiggs_{SM}$ & $\PZ^0\,\PZ^0$ &$\PHiggs_{SM}\,\PHiggs_{SM}$ & $\PZ^0\,\PZ^0$ \\ \hline \hline
$\sigma(\sqrt{S} = 1.5\,\TeV) [\picobarn]$ & $3.1\times10^{-4}$ & $0.02$  & $2.5\times10^{-5}$ & $0.06$ \\ \hline
$\sigma(\sqrt{S} = 3\,\TeV) [\picobarn]$  & $1.1\times10^{-3}$ & $0.06$  & $2.3\times10^{-4}$& 0.11 \\ \hline
\end{tabular}
\end{center}
\caption{\footnotesize{Cross-sections for the leading background
processes within the Standard Model, these are the SM Higgs boson
pair-production, $\APelectron\Pelectron \to
\PHiggs_{SM}\,\PHiggs_{SM} \,+ X$, and the $\PZ^0$ boson pair
production, $\APelectron\Pelectron \to Z^0 Z^0 \,+ X$. The
corresponding production rates are computed for $M_{\PHiggs_{SM}} =
(115,\,300)\,\GeV$ in order to account for phenomenologically
relevant scenarios. For a sufficiently heavy SM Higgs, resonant
production of $Z^0$-boson pairs occurs.}} \label{tab:sm}
\end{table}

%%%%%%%%%%%%%%%%%%%%%%%%%%%%%%%%%%%%%%%%%%%%%%%%%%%%%%%%%%%%%%%
\noindent \mysection{Conclusions} \label{sect:conclusions} We have
devoted this work to the study of the inclusive production of
Higgs-boson pairs in $\APelectron\Pelectron$ collisions (\ref{2HX}),
mainly triggered by the weak gauge boson fusion mechanism
(\ref{2HXF}) within the general 2HDM. We have found that the
cross-sections for some of these processes can be several orders of
magnitude larger than their MSSM counterparts. Moreover, in contrast
to the two-body final states (\ref{2H}), a tree-level analysis of
the gauge fusion mechanism could reveal the general 2HDM nature of
the Higgs bosons involved, if the enhancement properties that we
have explored effectively apply in the physical region of the
parameter space. This was shown to be the case also for the
previously considered processes (\ref{3H}) with three Higgs bosons
in the final state\,\cite{previous}. However, the inclusive 2HX
channels (\ref{2HXF}) could be by far the leading mechanism for
Higgs boson production at the characteristic TeV energies of the
planned ILC and CLIC colliders\,\cite{ILC-CLIC}. We find remarkable
opportunities whose threefold origin stems from: i) the sustained
(logarithmic) growing of the weak gauge boson fusion channels with
$\sqrt{S}$ at TeV energies; ii) the enhanced regime of the trilinear
(3H) Higgs boson couplings in the 2HDM; and iii) the possible
resonant (or near resonant) decay of an intermediate Higgs boson
into the final Higgs boson-pairs.

Phenomenologically interesting results are attained in large regions
of the 2HDM parameter space. In these domains, the maximum
cross-section of the inclusive production of Higgs boson pairs
(\ref{2HX}) can be far above (one to two orders of magnitude) the
exclusive triple-Higgs boson events (\ref{3H}). The possible sources
of background (coming from SM Higgs- and gauge-boson pair
production) are mostly negligible. In the case of type-II models,
which are closer to the MSSM Higgs sector, two simultaneously of the
inclusive processes (\ref{2HX}) can have cross-sections above $0.01$
pb (and one of them at the level of $1$ pb), therefore with
production rates at the level of $10^3-10^5$ events per
$100\,\invfb$ of integrated luminosity. Let us note that type-I
models may also lead to cross-sections of order of  $1$ pb in
particular channels, but here we may have up to four channels
(\ref{2HX}) simultaneously having sizeable cross-sections of
$0.01-0.1$ pb, thus amounting to production rates of $10^3-10^4$
events for the same segment of integrated luminosity. In contrast,
the corresponding MSSM maximum cross-sections are typically of the
order of $(0.1-1)$ fb, and hence some $100-1000$ times smaller
(similar to the fusion production of Higgs boson pairs in the SM).

Let us also mention that for type-II models (characterized by a
heavier spectrum of Higgs boson masses), the leading decay modes for
each of the Higgs bosons in a typical final state will be into heavy
quarks. For example, if we take the channel (\ref{2HX})
corresponding to the Higgs boson pair $\hzero\,\hzero$, each of the
neutral Higgs bosons will mainly decay as $\hzero\rightarrow
\Pbottom \APbottom$. In this region of parameter space, the
alternate Higgs boson decays into gauge bosons (such as
$\hzero\rightarrow \PWplus\PWminus,\PZ\PZ$) are either kinematically
forbidden or simply not dominant. In practice we would therefore
expect to see a sizable number of $4$-prong final states consisting
of highly energetic ($M_{\hzero}/2>50\,GeV$)
$\Pbottom\,\APbottom$-jets (or $\Ptop\,\APbottom$-jets from charged
Higgs decays from the $\PHiggsplus\,\PHiggsminus$ final state),
which should be clearly distinguishable in a linear
$\APelectron\Pelectron$-collider.

We have also demonstrated that by exploring $\APelectron\Pelectron$
collisions at even higher energies (say, up to the characteristic
$3$ TeV range expected for CLIC) the opportunities could still be
better. The most favored channel may then reach a cross-section at
the level of $5$ pb, and thus producing around half million events
in the same range of integrated luminosity. In general, this
upgrading should enable to perform accurate measurements of the
cross-sections of several channels (\ref{2HX}) and disentangle
enough correlations among the parameters of the model so as to be
able to pin down the corresponding Higgs boson masses and couplings
at high precision. Needless to say, a truly accurate analysis
demands the incorporation of quantum effects in the trilinear
interactions. Such study, however, goes far beyond the scope of the
present letter, whose main aim is only to show that clear signs of
new physics can be highlighted already from an analysis of Higgs
boson production through gauge boson fusion at leading order.
{Further investigation on these 2HX channels may also be of interest
in the context of the LHC. However, appropriate studies of the
distribution of the signal versus the background should be
undertaken in order to ascertain whether the dominant signatures of
the 2HX events (in the form of heavy-quark jets) could be
disentangled from the important QCD background inherent in the
physics of that collider.}

In summary, experiments at linear colliders such as the ILC and CLIC
can provide the cleanest signals of new physics and can be of
paramount importance as a high precision tool to underpine the most
sensitive building blocks of the gauge theories, and most
significantly the structure of the Higgs potential.

%%%%%%%%%%%%%%%%%%%%%%%%%%%%%%%%%%%%%%%%%%%%%%%%%%%%%%%%%%%%%%%
\vspace{0.2cm}
 \noindent
\textbf{Acknowledgments}\,\, We thank J. Guasch and W. Hollik for
discussions. RNH and DLV acknowledge the support of respective ESR
positions of the EU project RTN MRTN-CT-2006-035505 Heptools; DLV
also acknowledges the support of the MEC FPU grant AP2006-00357. The
work of JS has been supported in part by MEC and FEDER under project
FPA2007-66665 and by DURSI Generalitat de Catalunya under project
2005SGR00564. RNH thanks the Dept. d'Estructura i Constituents de la
Mat\`eria of the Univ. de Barcelona for the hospitality; DLV and JS
are grateful to the Theory Group at the Max-Planck Institut f\"ur
Physik in Munich for the hospitality and financial support. This
work was partially supported by the Spanish Consolider-Ingenio 2010
program CPAN CSD2007-00042.

%%%%%%%%%%%%%%%%%%%%%%%%%%%%%%%%%%%%%%%%%%%%%%%%%%%%%%%%%%%%%%%%%%%%%%%%
% shortcuts for handling the bibliography
%%%%%%%%%%%%%%%%%%%%%%%%%%%%%%%%%%%%%%%%%%%%%%%%%%%%%%%%%%%%%%%%%%%%%%%%%
%\newcommand{\JHEP}[3]{{\sl J. of High Energy Physics } {JHEP} {#1} (#2)  {#3}}
\newcommand{\JHEP}[3]{ {JHEP} {#1} (#2)  {#3}}
\newcommand{\NPB}[3]{{\sl Nucl. Phys. } {\bf B#1} (#2)  {#3}}
\newcommand{\NPPS}[3]{{\sl Nucl. Phys. Proc. Supp. } {\bf #1} (#2)  {#3}}
\newcommand{\PRD}[3]{{\sl Phys. Rev. } {\bf D#1} (#2)   {#3}}
\newcommand{\PLB}[3]{{\sl Phys. Lett. } {\bf B#1} (#2)  {#3}}
\newcommand{\EPJ}[3]{{\sl Eur. Phys. J } {\bf C#1} (#2)  {#3}}
\newcommand{\PR}[3]{{\sl Phys. Rep } {\bf #1} (#2)  {#3}}
\newcommand{\RMP}[3]{{\sl Rev. Mod. Phys. } {\bf #1} (#2)  {#3}}
\newcommand{\IJMP}[3]{{\sl Int. J. of Mod. Phys. } {\bf #1} (#2)  {#3}}
\newcommand{\PRL}[3]{{\sl Phys. Rev. Lett. } {\bf #1} (#2) {#3}}
\newcommand{\ZFP}[3]{{\sl Zeitsch. f. Physik } {\bf C#1} (#2)  {#3}}
\newcommand{\MPLA}[3]{{\sl Mod. Phys. Lett. } {\bf A#1} (#2) {#3}}
\newcommand{\JPG}[3]{{\sl J. Phys.} {\bf G#1} (#2)  {#3}}
\newcommand{\CPC}[3]{{\sl Comp. Phys. Comm.} {\bf G#1} (#2)  {#3}}
%%%%%%%%%%%%%%%%%%%%%%%%%%%%%%%%%%%%%%%%%%%%%%%%%%%%%%%%%%%%%%%%%%%%%%%%

\end{document}